\documentclass[preprint,amsmath,amssymb,superscriptaddress,prb,floatfix,footinbib]{revtex4-1}
\usepackage[]{graphicx}
\usepackage{tabularx}
\usepackage[usenames,dvipsnames]{color}
\usepackage{soul}
\usepackage{bm}
\usepackage{amsmath}
\usepackage{lineno}

\usepackage{times}
\usepackage{amsfonts}
\usepackage{mathrsfs}
\usepackage{graphicx}
\usepackage{dcolumn}
\usepackage{bm}
\usepackage{color}

\usepackage[colorlinks,bookmarks=false,citecolor=blue,linkcolor=red,urlcolor=blue]{hyperref}
\usepackage{multirow}
\usepackage{physics}

\begin{document}






\title{Dynamics and Resilience of the Charge Density Wave in a Bilayer Kagome Metal}

\author{Manuel Tuniz**}
\affiliation{Dipartimento di Fisica, Universita degli studi di Trieste, 34127, Trieste, Italy}

\author{Armando Consiglio**}
\affiliation{Institut f\"{u}r Theoretische Physik und Astrophysik and W\"{u}rzburg-Dresden Cluster of Excellence ct.qmat, Universit\"{a}t W\"{u}rzburg, 97074 W\"{u}rzburg, Germany}

\author{Denny Puntel}
\affiliation{Dipartimento di Fisica, Universita degli studi di Trieste, 34127, Trieste, Italy}

\author{Chiara Bigi}
\affiliation{School of Physics and Astronomy, University of St Andrews, St Andrews KY16 9SS, United Kingdom}


\author{Stefan Enzner}
\affiliation{Institut f\"{u}r Theoretische Physik und Astrophysik and W\"{u}rzburg-Dresden Cluster of Excellence ct.qmat, Universit\"{a}t W\"{u}rzburg, 97074 W\"{u}rzburg, Germany}

\author{Ganesh Pokharel}
\affiliation{Materials Department, University of California Santa Barbara, Santa Barbara, California 93106, USA}

\author{Pasquale Orgiani}
\affiliation{Istituto Officina dei Materiali, Consiglio Nazionale delle Ricerche, Trieste I-34149, Italy}

\author{Wibke Bronsch}
\affiliation{Elettra - Sincrotrone Trieste S.C.p.A., Strada Statale 14, km 163.5, Trieste, Italy}

\author{Fulvio Parmigiani}
\affiliation{Dipartimento di Fisica, Universita degli studi di Trieste, 34127, Trieste, Italy}
\affiliation{Elettra - Sincrotrone Trieste S.C.p.A., Strada Statale 14, km 163.5, Trieste, Italy}

\author{Vincent Polewczyk}
\affiliation{Istituto Officina dei Materiali, Consiglio Nazionale delle Ricerche, Trieste I-34149, Italy}

\author{Phil D. C. King}
\affiliation{School of Physics and Astronomy, University of St Andrews, St Andrews KY16 9SS, United Kingdom}

\author{Justin W. Wells}
\affiliation{Centre for Materials Science and Nanotechnology, University of Oslo (UiO), Oslo 0318, Norway.}

\author{Ilija Zeljkovic}
\affiliation{Department of Physics, Boston College, Chestnut Hill, MA 02467, USA}

\author{Pietro Carrara}
\affiliation{Dipartimento di Fisica, Università degli Studi di Milano, via Celoria 16, 20133 Milano, Italy}
\affiliation{Istituto Officina dei Materiali, Consiglio Nazionale delle Ricerche, Trieste I-34149, Italy}

\author{Giorgio Rossi}
\affiliation{Dipartimento di Fisica, Università degli Studi di Milano, via Celoria 16, 20133 Milano, Italy}
\affiliation{Istituto Officina dei Materiali, Consiglio Nazionale delle Ricerche, Trieste I-34149, Italy}

\author{Jun Fujii}
\affiliation{Istituto Officina dei Materiali, Consiglio Nazionale delle Ricerche, Trieste I-34149, Italy}
\author{Ivana Vobornik}
\affiliation{Istituto Officina dei Materiali, Consiglio Nazionale delle Ricerche, Trieste I-34149, Italy}
\author{Stephen D. Wilson}
\affiliation{Materials Department, University of California Santa Barbara, Santa Barbara, California 93106, USA}

\author{Ronny Thomale}
\affiliation{Institut f\"{u}r Theoretische Physik und Astrophysik and W\"{u}rzburg-Dresden Cluster of Excellence ct.qmat, Universit\"{a}t W\"{u}rzburg, 97074 W\"{u}rzburg, Germany}

\author{Tim Wehling}
\affiliation{Institute of Theoretical Physics, University of Hamburg, Notkestrasse 9, 22607 Hamburg, Germany}
\affiliation{The Hamburg Centre for Ultrafast Imaging, Luruper Chaussee 149, 22761, Hamburg, Germany}

\author{Giorgio Sangiovanni}
\affiliation{Institut f\"{u}r Theoretische Physik und Astrophysik and W\"{u}rzburg-Dresden Cluster of Excellence ct.qmat, Universit\"{a}t W\"{u}rzburg, 97074 W\"{u}rzburg, Germany}

\author{Giancarlo Panaccione}
\affiliation{Istituto Officina dei Materiali, Consiglio Nazionale delle Ricerche, Trieste I-34149, Italy}

\author{Federico Cilento}\email{federico.cilento@elettra.eu}
\affiliation{Elettra - Sincrotrone Trieste S.C.p.A., Strada Statale 14, km 163.5, Trieste, Italy}

\author{Domenico Di Sante}\email{domenico.disante@unibo.it}
\affiliation{Department of Physics and Astronomy, University of Bologna, 40127 Bologna, Italy}
\affiliation{Center for Computational Quantum Physics, Flatiron Institute, 162 5th Avenue, New York, NY 10010, USA}

\author{Federico Mazzola}\email{federico.mazzola@unive.it}
\affiliation{Department of Molecular Sciences and Nanosystems, Ca’ Foscari University of Venice, 30172 Venice, Italy}
\affiliation{Istituto Officina dei Materiali, Consiglio Nazionale delle Ricerche, Trieste I-34149, Italy}

\date{\today}

\begin{abstract}
\bf{Long-range electronic order descending from a metallic parent state constitutes a rich playground to study the intricate interplay of structural and electronic degrees of freedom \cite{Neupert_2022,Arachchige_2022,Wang_2021a, Heumen_2021, Luo_2022}. With dispersive and correlation features as multifold as topological Dirac-like itinerant states, van-Hove singularities, correlated flat bands, and magnetic transitions at low temperature, kagome metals are located in the most interesting regime where both phonon and electronically mediated couplings are significant \cite{Kang_2020, Kang_2020b, Lin_2018, Ye_2019,Pokharel_2021,Hu_2022}. Several of these systems undergo a charge density wave (CDW) transition, and the van-Hove singularities, which are intrinsic to the kagome tiling, have been conjectured to play a key role in mediating such an instability. However, to date, the origin and the main driving force behind this charge order is elusive. Here, we use the topological bilayer kagome metal ScV$_6$Sn$_6$ as a platform to investigate this puzzling problem, since it features both kagome-derived nested Fermi surface and van-Hove singularities near the Fermi level, and a CDW phase that affects the susceptibility, the neutron scattering, and the specific heat \cite{Arachchige_2022}, similarly to the siblings $A$V$_3$Sb$_5$ ($A$ = K, Rb, Cs) \cite{Neupert_2022, Kang_2023} and FeGe \cite{Teng_2022}. We report on our findings from high—resolution angle-resolved photoemission, density functional theory, and time-resolved optical spectroscopy to unveil the dynamics of its CDW phase. We identify the structural degrees of freedom to play a fundamental role in the stabilization of charge order. Along with a comprehensive analysis of the subdominant impact from electronic correlations, we find ScV$_6$Sn$_6$ to feature an instance of charge density wave order that predominantly originates from phonons. As we shed light on the emergent phonon profile in the low-temperature ordered regime, our findings pave the way for a deeper understanding of ordering phenomena in all CDW kagome metals.}
\end{abstract}

\maketitle
\begin{figure*}[!h]
\centering
\includegraphics[width=\textwidth,angle=0,clip=true]{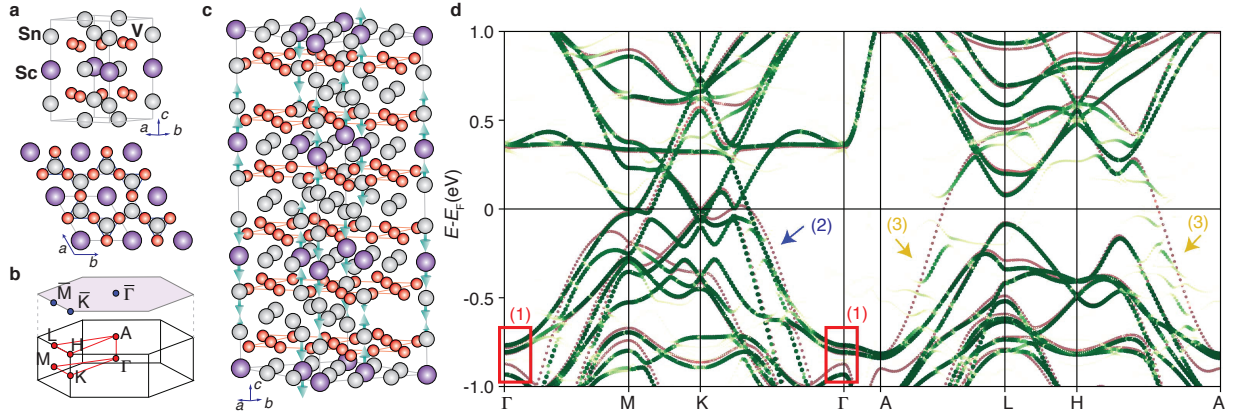}
\caption{{\bf Crystal and calculated electronic structure of ScV$_6$Sn$_6$.} {\bf a} Crystal structure without CDW. The top-view corresponds to a 2$\times$2 cell. {\bf b} Brillouin Zone with high-symmetry points and directions. {\bf c} Crystal structure in the CDW phase with the out-of-plane distortions indicated by the arrows. The length of the arrow is proportional to the displacement. {\bf d} Electronic structure of ScV$_6$Sn$_6$ without (red lines) and with CDW (green color) along the high symmetry directions. For visual clarity, the pristine bands have been translated in energy by +0.029 eV. In this way the van-Hove singularities and the $d_{z^2}$ flat band of the CDW and pristine systems are aligned. It is indeed expected that a structural distortion leads to a tuning in the electronic band structure, especially around E$_F$ \cite{PhysRevB.105.165146}. A similar but disentangled figure can be found in the Supplementary Information. The main changes are here indicated: at $\Gamma$ (red box, label '1') the CDW induces an increment in the bands separation. Along both $\Gamma$-K and $\Gamma$-M the bands appear to have slightly different k-loci (blue arrow and label '2'). The main changes are observed along the A-H and A-L directions (yellow arrows and label '3') and are characterized by a marked Sn $p_z$-states}.
\label{fig1}
\end{figure*}
\begin{figure*}[!h]
\centering
\includegraphics[width=\textwidth,angle=0,clip=true]{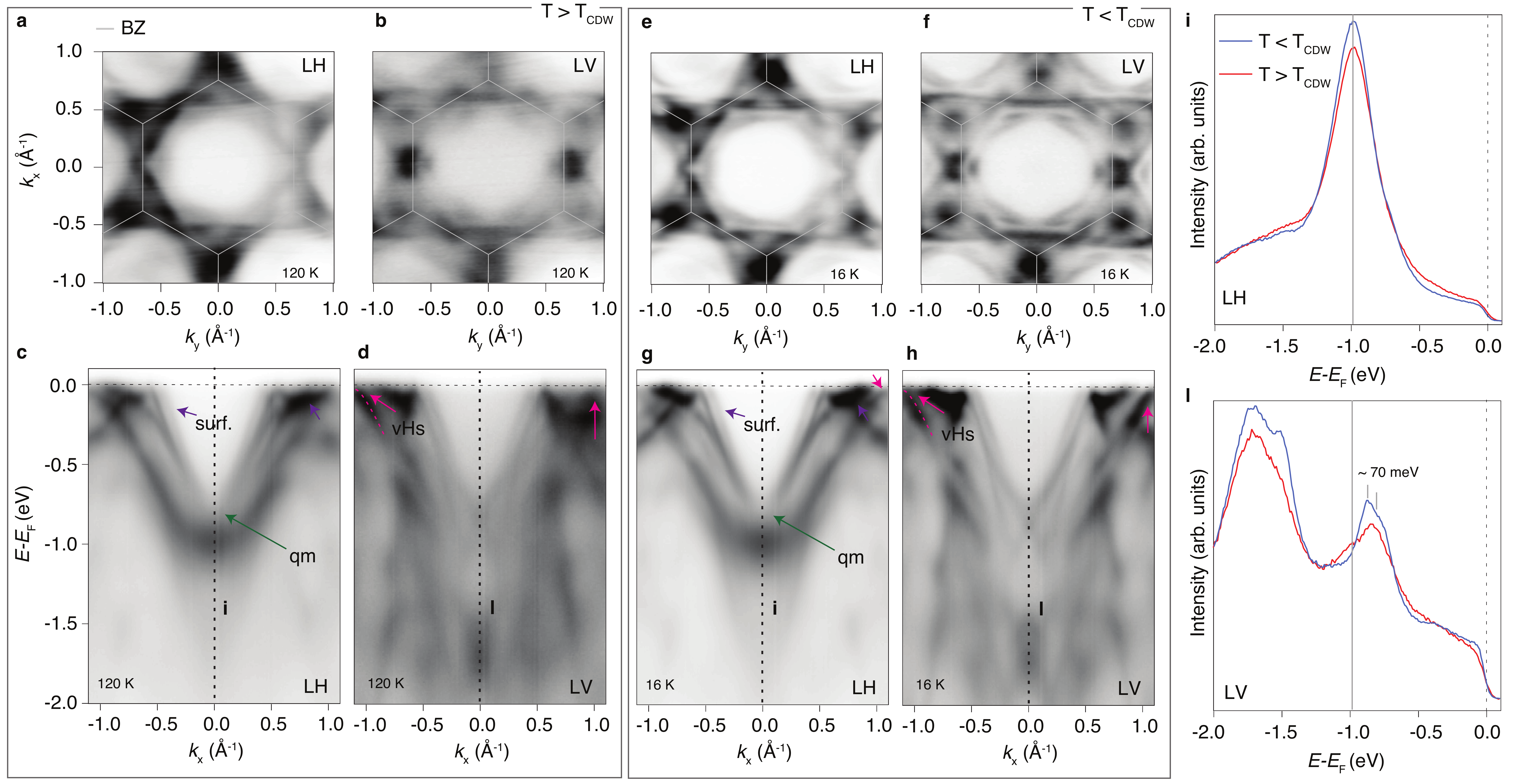}
\caption{{\bf Measured electronic structure of ScV$_6$Sn$_6$ across the CDW critical temperature.} Fermi surface of ScV$_6$Sn$_6$ collected above $T_{CDW}$ by using linear {\bf a} horizontal and {\bf b} vertical light polarizations and {\bf c-d} corresponding energy versus momentum dispersion along the $\Gamma$-K-M direction of the Brillouin zone. Fermi surface of ScV$_6$Sn$_6$ collected below $T_{CDW}$ by using linear {\bf e} horizontal and {\bf f} vertical light polarizations and {\bf g-h} corresponding energy versus momentum dispersion along the $\Gamma$-K-M direction of the Brillouin zone. {\bf i} Energy distribution curve (EDC) extracted at the $\Gamma$-point for LH and {\bf l} LV polarization, across the transition temperature. LV reveals more spectral details, including a small doublet.}
\label{fig2}
\end{figure*}
\begin{figure*}[!h]
\centering
\includegraphics[width=\textwidth,angle=0,clip=true]{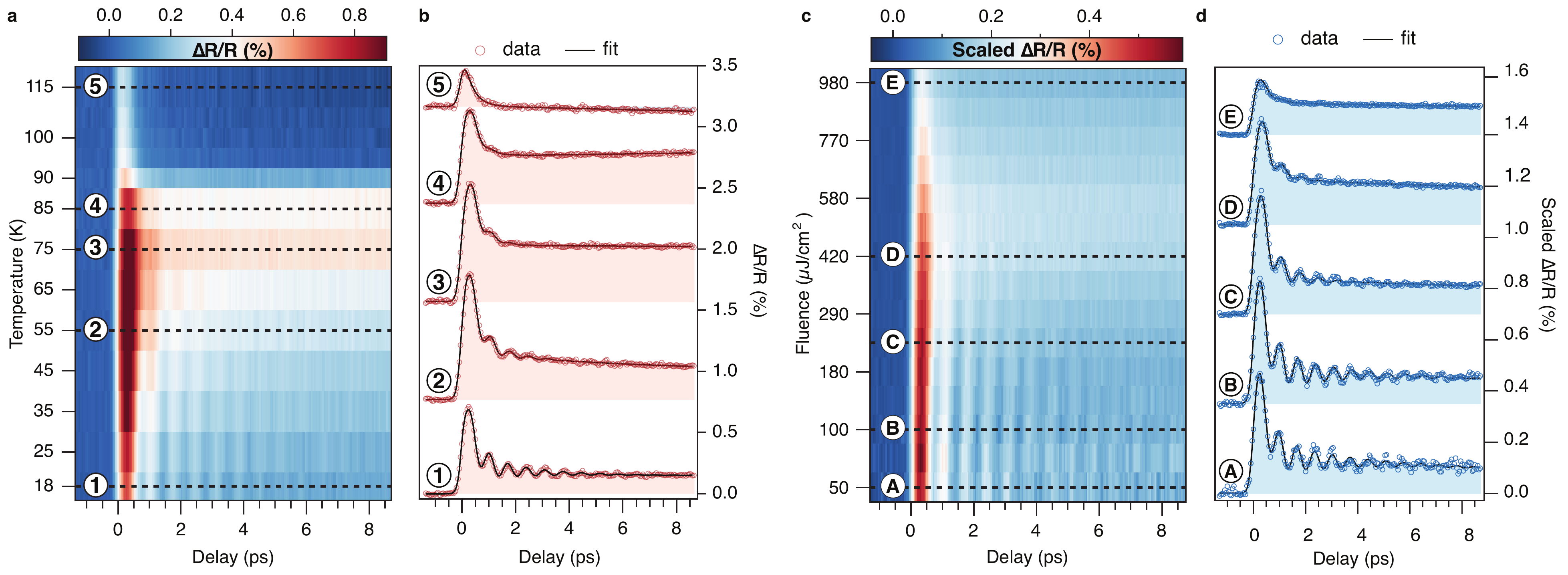}
\caption{{\bf Temperature and fluence dependent-dynamics of ScV$_6$Sn$_6$.} {\bf a} Evolution of $\Delta$R/R as a function of the temperature in the temperature range across $T_{CDW}$, showing the signatures of the phase transition around 90~K. In the low temperature CDW phase a clear oscillatory response is observed. {\bf b} $\Delta$R/R profiles extracted from {\bf a} at selected temperatures, showing the evolution of the dynamics across the phase transition. {\bf c} Evolution of the scaled $\Delta$R/R as a function of the fluence across $T_{CDW}$. The traces are scaled to the fluence to ease the visualization upon approaching the photoinduced phase transition. {\bf d} Scaled $\Delta$R/R profiles extracted from {\bf c} at selected fluences.}
\label{fig3}
\end{figure*}
\begin{figure*}[!h]
\centering
\includegraphics[width=\textwidth,angle=0,clip=true]{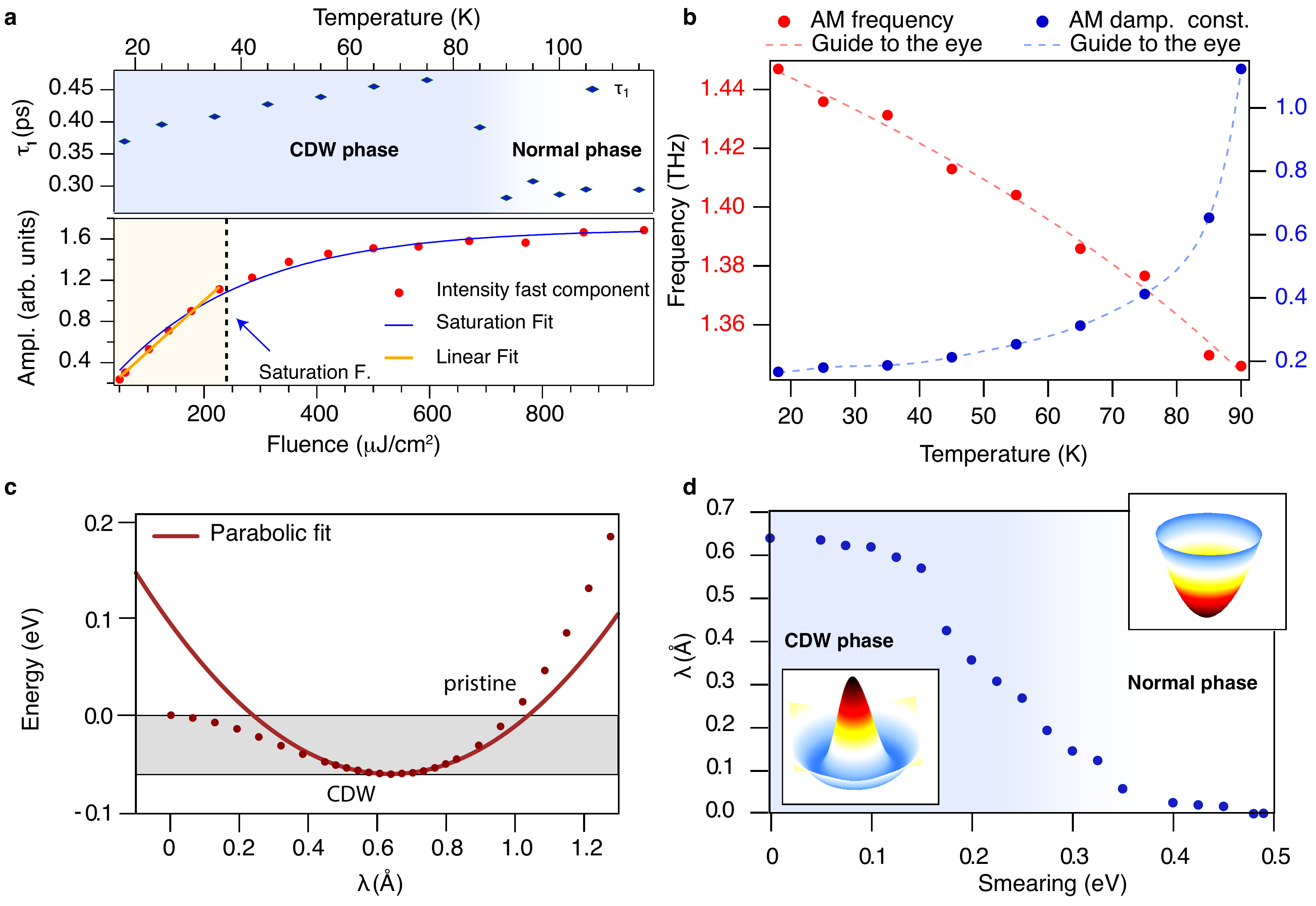}
\caption{{\bf Evolution of significant fit parameters.} 
{\bf a} Evolution of the lifetime of the first fast decay ( $\tau_1$) as a function of the temperature (top panel) and evolution of the amplitude of the fast peak as function of fluence (bottom panel). {\bf b} Evolution of the frequency and the damping constant of the amplitude mode as a function of the temperature. {\bf c} Total energy  of the ScV$_6$Sn$_6$ constant-volume superlattice while interpolating  between the pristine, the CDW and the exaggerated CDW phases. $\lambda =  0.00$ \AA ~ means that the displacement vector has a vanishing norm, i.e.  we are in the reference starting point (the pristine phase).  When $\lambda = 0.64$ \AA, the norm of the displacement vector is  $\norm{\Vec{e}} = \lambda$, hence we are in the CDW phase. When further increasing the norm of the displacement vector, we are describing an exaggerated CDW phase. The parabola represented by a  continuous line is the quadratic fit of the energy profile, around the  minimum. The quadratic coefficient is related to the spring constant $k$ and frequency $\omega$ of the CDW amplitude mode (more details in the Methods section). {\bf d} Norm of the  351-dimensional displacement vector $\Vec{e}$ among the CDW and  pristine structures, as a function of the electronic (gaussian) smearing $\sigma$ used  during the constant-volume ionic relaxation.}
\label{fig4}
\end{figure*}

** These authors contributed equally\\
Bilayer kagome metals are an emerging class of correlated topological compounds with chemical formula  $R$V$_6$Sn$_6$ ($R$: rare earth - hereafter \textcolor{black}{also} dubbed \textcolor{black}{as} the 166 family), featuring unconventional topological phases and collective magnetic excitations at low temperature \cite{Pokharel_2021, Hu_2022, Arachchige_2022, Peng_2021, Ishikawa_2021, Rosenberg_2022, Lee_2022, Chen_2021}. Similarly to the sister compounds $A$V$_3$Sb$_5$ ($A$: alkali atoms - hereafter \textcolor{black}{also nicknamed as} the 135 family), the electronic structure of $R$V$_6$Sn$_6$ is a rich platform for the emergence of unconventional phase diagrams, boasting the simultaneous presence of itinerant Dirac electrons, non-trivial flat bands, and van-Hove singularities at the Fermi energy \cite{Kang_2020, Kang_2020b, Lin_2018, Ye_2019,Guo2009,Mazin2014,Kiesel2013,DiSante2020,Neupert_2022}. Among the 166 family, ScV$_6$Sn$_6$ exhibits a unique three-dimensional charge density wave (CDW) transition, with a motif dissimilar to those observed in other kagome compounds \cite{Kang_2023}. Here, the CDW phase provokes a weak atomic lattice displacement, which might generate unconventional phenomena, such as metal-insulator transitions of fermions, vortexes at the surface, chiral currents, and bond orders \cite{Arachchige_2022, Guo_2022, Nishimoto_2010, Kiesel_2013, Denner_2021, Feng_2021, Park_2021, Jiang_2021, Liang_2021}. The major goal of contemporary research thus is to reveal the underlying mechanism of the charge order in ScV$_6$Sn$_6$ and the specific role of structural and electronic degrees of freedom, the interdependence of whom has so far remained uncharted territory.\\
Previous works have highlighted important differences between the CDW in ScV$_6$Sn$_6$, $A$V$_3$Sb$_5$, and FeGe \cite{Teng_2022}. In $A$V$_3$Sb$_5$, the charge order has been thought to be connected to the nesting of the Fermi surface, exemplified by the van-Hove singularity electron filling \cite{Wu2021, Kang2022, Ortiz2020, Kang_2023}. In these systems, the charge order has been conjectured to be linked to the emergence of superconductivity and stripe orders \cite{Yu_2021, Liang_2021, Du_2021, Oey_2022, Li_2021a, Wang_2021b, Zhao_2021}. In addition, the atomic displacement following the onset of the CDW involves mostly the V atoms and their in-plane positions. ScV$_6$Sn$_6$, surprisingly, does not undergo any superconducting transition at low temperature and the motif of the CDW, generated by the ordering vector $Q=(1/3,1/3,1/3)$, does not give rise to either the Star of David or trihexagonal arrangement, as for the $A$V$_3$Sb$_5$ compounds, for which the ordering vectors are $Q=(1/2,1/2,1/4)$ and $Q=(1/2,1/2,1/2)$ \cite{Ortiz_2021, Ortiz2020, Nie_2022}. This is ascribed to the involvement of Sn and Sc atoms in the lattice modulation. The differences between these two systems, are also reflected in a much more rapid CDW suppression for ScV$_6$Sn$_6$ under the application of pressures \cite{Zhang_2022}, offering an increased ease of tuneability compared to $A$V$_3$Sb$_5$.\\
In CDW materials, the simultaneous occurrence of a lattice displacement and a change in the electronic structure, often associated with the opening of gaps in the electronic spectrum challenges the understanding of the main driving force behind the charge order and limits the access to possible consequent or concomitant cascade of collective phases and their control. While there is compelling evidence of CDW in kagome systems, its microscopic origin and dynamics are still debated. We stress that the understanding of such origin and dynamics is a task of paramount importance given the peculiar electronic structure of kagome with van-Hove singularities at the Fermi level and the presence of a nested Fermi surface, which hints at an energy gain upon a gap opening as an underlying driver. For 166, which has also a topological nature, it becomes crucial to understand the dynamics of the CDW, to uncover not only the relationship between subsequent ordered phases, but to elucidate the interplay between phase transition and electron spins \cite{Hu_2022, Yin_2022, Sun_2022, Tanaka_2020a, Ortiz2020, NagaosaPRB2019}. In the present study, we used time-resolved optical spectroscopy (TR-OS) to unveil the dynamics of the CDW in the 166 bilayer kagome ScV$_6$Sn$_6$. Specifically, by using TR-OS, we disclose the energy scales involved in the charge order, which points towards a major robustness of the lattice degree of freedom for the observed transition.\\
First, we describe the electronic properties of ScV$_6$Sn$_6$, then we will show by using ARPES, DFT, and TR-OS how the electronic and structural degrees of freedom are affected by the CDW order. Along with the crystal structure (both with and without CDW) and the Brillouin zone (Figs.\ref{fig1}~{\bf a}-{\bf b}), we show the expected electronic structure from density functional theory (DFT), above and below the transition temperature, in Fig.\ref{fig1}~{\bf c} (red color without CDW, green shades with CDW). In both CDW and pristine cases, the Dirac-like states and the van-Hove singularities are similar and identifiable, with negligible variation in their electronic dispersion along the $\Gamma$-K-M-$\Gamma$ path. According to the calculations, the van-Hove singularities are located around the Fermi level. It is thus noteworthy that large changes at E$_F$ are not seen upon inclusion of the CDW nested systems \cite{Whangbo_1991, Carpinelli_1996,Tam_2022, Bosak_2021}. Along the $\Gamma$-K-M-$\Gamma$ high-symmetry direction, the main changes occur at the $\Gamma$ point within the energy window [-0.75, -1]~eV. Below the transition temperature, the separations between the bands increase by approximately 30 meV compared to the electronic structure without CDW (see red box in Fig.\ref{fig1}~{\bf c}). Other changes along the same direction, even if smaller, are indicated by the blue arrows (labeled as '2'). If in the plane that comprises the $\Gamma$ point, the changes are rather subtle, our DFT calculation reveals that the CDW has its major effect along the A-H and A-L directions (see yellow arrows labeled as '3'). There, we notice that the charge order opens several energy gaps along the entire $k$-space path. Nonetheless, we also notice that the spectral weight immediately fades away from the gap openings. In addition, the spectral intensity in the CDW phase is still located around the poles of the electronic structure of the bands above $T_{CDW}$. As a result, the gaps themselves are expected to be challenging to detect experimentally given the vicinity to the poles, the $k_z$ broadening, and the experimental resolution ($\sim$ 15 meV, that is larger than the gap value itself) \cite{Sunko_2020}.\\
Similarly to other 166 materials, as shown in Fig. \ref{fig2}, ScV$_6$Sn$_6$ shows the presence of surface states, after the ultrahigh vacuum (UHV) cleave (see Methods for details about the samples' preparation). These surface states have been extensively discussed for other kagome systems \cite{Kang2022, Kang_2020, Kang_2020b} and members of the 166 family \cite{Pokharel_2021,Peng_2021, Ishikawa_2021, Rosenberg_2022, Hu_2022}. While they offer an interesting platform for exploiting spin-momentum locking because of their non-trivial topology \cite{Hu_2022, Li_2014, Li_2018, Jiang_2016, Luo_2017}, they could challenge the clear visualization of small changes occurring in the bulk electronic structure as a consequence of the charge order, because they could obfuscate subtle spectroscopic features. Thus, to exclude this avenue, we measured the electronic structure above and below the transition temperature $T_{CDW}$ and along the $\Gamma$-K-M-$\Gamma$ high symmetry direction, with and without surface states such that one can better distinguish the bulk from the surface-derived signal (see Fig.\ref{fig2} for the clean surface case with both light polarizations and Supplementary Information Figs. 1 and 2 for data where the surface states have been suppressed and how). Along the selected high-symmetry path, we have the most favorable photoemission matrix elements and the intensity of the bands is the most prominent. Notably, no remarkable differences are observed across $T_{CDW}$, besides a constant thermal broadening which affects the spectra collected above $T_{CDW}$.\\
To still check possible effects of the charge order, from the ARPES spectra collected along the $\Gamma$-K-M-$\Gamma$ path, we extracted energy-distribution curves (EDCs) across the $\Gamma$ point, where we expect the strongest changes induced by the CDW (Fig.\ref{fig2}~{\bf i}-{\bf l} and red box of Fig.\ref{fig1}~{\bf c}). By using linear-horizontal (LH) polarized light, it is hardly possible to see the relevant states affected by the charge order (see Fig.\ref{fig2}~{\bf i}). By using linear-vertical (LV), instead, the missing spectral weight is recovered. In this condition, we resolve a 70~meV splitting at approximately $0.8$~eV of binding energy. Such a  splitting might be consistent with the calculated increase in the energy separation produced by the CDW, however, we cannot exclude that this might be an effect of the band sharpening due to the lower temperature (see Fig.\ref{fig2}~{\bf l}). Despite having weaker intensity due to photoemission matrix elements, we also collected ARPES spectra along the A-L and A-H high-symmetry directions (shown in Supplementary Information Figs. 3, 4, and 5). Along these paths, the DFT analysis predicts the strongest changes in the electronic structure. Unfortunately, the changes from above to below $T_{CDW}$ are very small and not well-resolved experimentally. This is likely due, as anticipated above, to the combination of $k_z$ broadening, finite experimental resolutions, small energy values of the gaps, and expected spectral weight concentrated along the poles of the bands above the CDW transition.\\
The ARPES data depict a situation where most of the electronic spectral weight is only weakly affected by the CDW order in the proximity of the Fermi level (within the experimental resolutions). This is in line with DFT only showing a partial gap opening, which involves only certain bands but not the majority of the V-derived spectral weight. In addition, the partial gap calculated in the electronic structure agrees well with recent optical measurements in the static regime performed for ScV$_6$Sn$_6$, which show a sharp first order transition without the development of full gaps in the optical spectra \cite{Arachchige_2022}. Importantly, this behaviour describes a rather rare phenomenology, which is different from the most commonly observed second-order density condensation gaps, found in systems which undergo a CDW transition. In other words, the inability to spectroscopically resolve changes in the Fermi surfaces across the transition temperature makes ScV$_6$Sn$_6$ reminiscent of the CDW behaviour discovered for IrTe$_2$ , TaTe$_2$, and BaNi$_2$As$_2$ \cite{Li_2014b, Hu_2022, Fang_2013}. This aspect is intimately related to the underlying electronic structure of the compound; the vanadium states indeed contribute in a significant way to the density of states around $E_F$, while the corresponding atoms are not influenced by the CDW displacement. Conversely, tin and scandium states are not predominant at $E_F$ while the corresponding atoms' positions are particularly influenced by the CDW transition. Additionally, ScV$_6$Sn$_6$ shows a very peculiar ordered phase, with a marked peak in the specific heat and an increased metallic character below the transition \cite{Arachchige_2022}. However, the origin of the charge order is still an open question. It is then crucial to separate the time- and energy-scales of the electronic and lattice degrees of freedom and to determine their robustness and involvement in the CDW order.\\
To this aim, we use TR-OS and we measure the time-resolved reflectivity dynamics of ScV$_6$Sn$_6$. Fig.\ref{fig3}~{\bf a} presents the temperature (T) dependence of the photoinduced reflectivity transients, $\Delta R/R$, recorded upon increasing the temperature (T from approximately 18~K; the fluence was F $\approx$ 0.2~mJ/cm$^{2}$). The $\Delta R/R$ can be described by two features: an incoherent (non-oscillating) response and a coherent oscillatory behaviour. In our data, clear oscillations in the $\Delta R/R$, showing a strong T-dependence, are detected up to the transition temperature $T_{CDW}$ (Fig.\ref{fig3}~{\bf a-b}). The incoherent (non-oscillating) response consists of two decays, a fast and a slow one, with characteristic times $\tau_1$ and $\tau_2$, respectively (see Supplementary Information for details about the data analysis and their extraction) that appear in the data as a fast peak and a slowly-decaying component. Similarly to the coherent response, the incoherent one is also strongly T-dependent and it shows the indication of a phase transition around T $\approx$ 90~K (see Fig.\ref{fig4}~{\bf a}), which matches well with the nominal $T_{CDW}$ \cite{Arachchige_2022}. Above $T_{CDW}$, the $\Delta R/R$ becomes smaller and, after the fast exponential decay, it changes sign becoming negative (see trace '5' of Fig.\ref{fig3}~{\bf b}).\\
In general, below the critical CDW temperature, the pump pulse induces an abrupt change in the reflectivity, followed by a fast decay within a picosecond and a slower recovery on a timescale of the order $\approx$ 10 ps \cite{Tomeljak_2009,Yusupov_2008,Pokharel2022,Vorobeva_2011,Ravnik_2018}. The first, which slows down critically upon approaching the transition temperature, is attributed to the re-opening of the CDW gap, while the second one, as confirmed by detailed studies of the dynamics as a function of the excitation fluence and applied external electric field, is attributed to a second stage of the CDW recovery \cite{Eichberger2010,TOMELJAK2009_physB,Huber_2014,Storeck_2020}. 
 In ScV$_6$Sn$_6$, the T-evolution of the incoherent part of the $\Delta R/R$ shows a rather peculiar behavior: in contrast to the well-studied one dimensional CDW systems, the lifetime of the first fast decay $\tau_1$ increases linearly with the temperature and shows a rapid drop at $T_{CDW}$ (Fig.\ref{fig4}~{\bf a}). Remarkably, no divergence of the lifetime $\tau_1$ is observed during the phase transition. This peculiar behavior, in combination with our ARPES measurements and DFT analysis, suggests an unusual origin for the CDW phase in this material, where no gaps are observed in the near Fermi level region. Indeed, a divergence of the lifetime $\tau_1$ is usually linked to the closing of the electronic gap that develops in conventional CDW transitions. This is also different from what has been observed in the kagome metal CsV$_3$Sb$_5$, where TR-OS measurements have highlighted the divergence of the lifetime of the fast decay process during the onset of the charge order \cite{Wang_2021b}.\\
Interestingly, from the fits to the data (see Supplementary Information for more details), the frequency of the oscillatory mode and the damping show a pronounced temperature dependence, as we show in Fig.\ref{fig4}~{\bf b}. In particular, at low temperature the mode is characterized by a frequency of $\sim$ 1.45 THz, in excellent agreement with our ab-initio estimate of 1.42 THz from the quadratic fit of the DFT total energy around the minimum of CDW ordered structure (see Fig.\ref{fig4}~{\bf c} and Methods for more details), and we observe a $\approx$ 6 $\%$ softening of the mode frequency and an exponential increase of the damping constant, when the temperature of the system approaches the critical temperature of the CDW transition. Such a temperature dependence has been observed in many CDW systems and it constitutes the fingerprint of an amplitude mode (AM) of the CDW phase \cite{Demsar_1999,Chen_2017,Yusupov2010,Yoshikawa2021,Vorobeva_2011,Ravnik_2018}. This observation confirms that the coherent response is not due to a trivial phonon, but rather to the excitation of the AM of the CDW phase, confirming the onset of CDW physics in ScV$_6$Sn$_6$. We also note that no fingerprints of an AM have been hitherto detected in CsV$_3$Sb$_5$ by previous reflectivity studies \cite{Wang_2021b}.\\
By varying the excitation density (fluence), additional information about the nature of the phase transition can be obtained. The melting of electronic modulation is generally achieved on a timescale much faster than the characteristic timescale related to the relaxation of the periodic lattice modulation, which is given by the period of the characteristic amplitude modes of the system. This results in the disentanglement of electronic and lattice degrees of freedom on the sub-picosecond time scale, with a subsequent rapid recovery of the electronic degrees of freedom \cite{Schaefer2013,Yusupov2010,Vorobeva_2011,Stojchevska_2011}. Additionally, the characteristic energy required to fully drive the phase transition, i.e. quenching the periodic lattice distortion and the electronic degrees of freedom, is that one which leads to the disappearance of the (oscillatory) amplitude mode \cite{Pokharel2022,Schaefer2013}. This provides a means to quantify the strength of the lattice degree of freedom compared to the electronic one. We report this in Fig. \ref{fig3}~{\bf c}, where the evolution of the scaled $\Delta R/R$ signal as a function of the fluence is shown, in a range from $\sim50$ to $\sim 1000~\mathrm{\mu}$J/cm$^{2}$. The traces are scaled to the fluence to better visualize the qualitative behavior of the response upon approaching the photoinduced phase transition. Note that the sample is kept at a temperature well-below $T_{CDW}$. In the low perturbation regime, the signal scales linearly with excitation density (traces A, B, C in Fig. \ref{fig3}~{\bf d}), while upon approaching the photoinduced phase transition the $\Delta R/R$ signal related to CDW order shows saturation \cite{Schaefer2013,Pokharel2022}. 
The maximum change that can be induced in the signal is generally achieved when the electronic degree of freedom associated to the CDW order is collapsed. 
We find that the fast electronic component with relaxation time $\tau_1$ saturates at a fluence, where the oscillations linked to AM involving the lattice degrees of freedom persist  \cite{Schaefer2013,Tomeljak_2009, Pokharel_2022}. By progressively increasing the excitation fluence, the oscillations in the $\Delta R/R$ due to the AM get also suppressed, with an exponential increase of the damping constant and a softening of the frequency of the mode. Remarkably, our measurements highlight the robustness of the lattice reconstruction of the CDW in this material. Indeed, even at high excitation fluences ($\sim700 \ \mu$J/cm$^{2}$), the oscillations due to the AM are still detected. This behavior is again markedly different from what has been observed in the sister compound CsV$_3$Sb$_5$, where already at a pump fluence as small as $\sim55 \ \mu$J/cm$^{2}$, the CDW order is melted \cite{Wang_2021a}. In our study, we show that the oscillation disappears completely only at a fluence of $\sim1000 \ \mu$J/cm$^{2}$, meaning that the lattice order can be melted, although at a fluence larger than the one required to saturate the purely electronic response at around $ \approx$ 240 $\mu$J/cm$^{2}$. The electronic states that maximally couple to this CDW span an energy window around the Fermi level of about 0.25 eV, because smearing over that window greatly affects the AM, as we show in Fig. \ref{fig4}~{\bf d}. Indeed, taking into account the electronic specific heat of ScV$_6$Sn$_6$, an absorbed fluence of $ \approx$ 240 $\mu$J/cm$^{2}$ leads to an increase of the electronic temperature of $\approx$ 2100~K, which is in reasonable agreement with the smearing obtained by DFT calculations (See Fig. \ref{fig4}~d). Finally, it is only when the average lattice temperature exceeds the CDW critical temperature because of the average power deposited by the pump pulse, that one might expect to detect the transient signal of the high-temperature phase. We estimate the pump-induced average heating to be in a $\Delta T\approx$ 20~K at a fluence of $\approx$ 1 mJ/cm$^{2}$. Hence, in order to reach $T_{CDW}$, very large fluences are required, which are not attainable because sample damage would occurr.\\
In conclusion, by using the combination of experimental techniques, i.e. ARPES and TR-OS, as well as theoretical calculations, we disclose the dynamics of the unusual CDW transition in ScV$_6$Sn$_6$ and we simultaneously demonstrate marginal energy gaps at the Fermi level. 
In particular, temperature dependent TR-OS reveals the hallmarks of the CDW. In addition, our fluence dependent study disentangles the electronic and structural degrees of freedom, showing persisting AM oscillation up to unusually high values of fluence, corroborating the robustness of the lattice distortions compared to the sudden disruption of the purely electronic component. Such dynamics 
is in favor of a CDW with the lattice degrees of freedom and electron-phonon interaction playing the dominant role. We believe that our comprehensive study sheds light onto the collective modes emerging in the low-temperature phase of ScV$_6$Sn$_6$. It paves the way for the understanding of unconventional many-body phases in correlated kagome metals, which all share a similarly nested Fermi surface, and deepens our understanding of cooperative lattice and electronic symmetry breaking at this new frontier. Therefore, we think that our discovery is the crucial step to understand the origin and drive of the CDW also in the archetypal family AV$_3$Sb$_5$.\\
\noindent{\bf Methods}\\
\noindent{\bf Experimental details --}
Single crystals of ScV$_6$Sn$_6$ were grown using a flux-based growth technique as reported in the reference \cite{PhysRevMaterials.6.104202}. Sc (chunk, 99.9$\%$), V (pieces, 99.7$\%$), and Sn (shot, 99.99$\%$) were loaded inside an alumina crucible with the molar ratio of 1:6:20 and then heated at 1125 $^{\circ}$C for 12 h. Then, the mixture was slowly cooled to 780 $^{\circ}$C at a rate of 2 $^{\circ}$C/h. Thin plate-like single crystals were separated from the excess Sn-flux via centrifuging at 780 $^{\circ}$C. The samples were cleaved in ultrahigh vacuum (UHV) at the pressure of 1$\times$10$^{-10}$ mbar. The ARPES data were acquired at the APE-LE end station (Trieste) using a VLEED-DA30 hemispherical analyzer. The energy and momentum resolutions were better than 12 meV and 0.02 \AA$^{-1}$, respectively. The temperature of the measurements was kept constant throughout the data acquisitions (16 K and 120 K, below and above $T_{CDW}$ respectively). Both linear polarizations were used to collect the data from the APE undulator of the synchrotron radiation source ELETTRA (Trieste). The photon energy used for the ARPES data was 75~eV. This choice was such that the spectra intensity was the most prominent, especially near the region of the quadratic minimum of the electronic structure, where the major changes with the CDW are expected.\\
Time-resolved reflectivity experiments were performed at the T-ReX laboratory (FERMI, in Trieste) with a probe photon energy of $\approx$ 0.95 eV (1300 nm). The measurements were performed using a Ti:sapphire femtosecond (fs) laser system, delivering, at a repetition rate of 250 kHz, $\approx 50$ fs light pulses at a wavelength of 800 nm (1.55 eV). The single color probe measurements were performed at a probe wavelength of 1300 nm, obtained by filtering a broadband (0.8–2.3 eV) supercontinuum probe beam, generated using a sapphire window.\\
\noindent{\bf Theoretical details --} DFT calculations have been performed using both Quantum Espresso and VASP packages. Phonon calculations are based on density functional perturbation theory, as implemented in Quantum Espresso \cite{doi:10.1063/5.0005082, Giannozzi_2009, Giannozzi_2017}. Exchange and correlation effects were included with the generalized gradient approximation (GGA) using the Perdew-Burke-Ernzerhof (PBE) functional \cite{PhysRevLett.77.3865}; the pseudopotentials are norm-conserving and scalar relativistic, containing core corrections \cite{PhysRevB.88.085117}. Self-consistent calculations and ionic relaxation  of the unit cell have been performed with a 9$\times$9$\times$6 $k$-grid; convergence threshold for both ionic minimization and electronic self-consistency are set to be $10^{-15}$ Ry. The kinetic energy cutoff for the wavefunctions is equal to 90 Ry. A ordinary Gaussian spreading of 0.006 Ry has been used. Once the structure was properly at equilibrium, with vanishing forces acting on each atom, we proceeded with the actual phonon calculations. The dynamical matrices have been obtained and computed for a $q$-grid of 3$\times$3$\times$6, with a self-consistency threshold of $10^{-15}$ Ry. These dynamical matrices are consequently Fourier-transformed to obtain the inter-atomic force constants (IFC) in real space; three translational acoustic sum rules have been imposed, via correction of the IFC. Finally, both phonon dispersion and phonon density of states are obtained. In particular, for the density of states calculations (see results in the Supplementary Information) we used 30$\times$30$\times$30 $q$-points. The presence of imaginary phonon modes has been verified also for a 6$\times$6$\times$9 and a 9$\times$9$\times$12 $q$-grid, during the dynamical matrices' calculation.
The remaining DFT calculations have been performed with VASP. To study the dependence of the CDW phase with respect to the smearing we used a plane-wave cutoff of 500 eV and a 9$\times$9$\times$9 $\Gamma$-centered $k$-mesh. Ionic relaxations have been performed with a constant volume. The relaxations of the ionic and electronic degrees of freedom were considered converged respectively below a threshold of $10^{-5}$ eV and $10^{-10}$ eV. Subsequently, for each smearing value, we computed the norm of the 351-dimensional (117 atoms $\times$3) displacement vector among the CDW and the pristine phases. Spin-orbit coupling has not been included for this set of calculations.
To study the structural interpolation among the CDW and pristine systems, leading to the double-well potential profile, we computed the 351-dimensional displacement vector $\Vec{e}$ between the two configurations. This has a norm of $\sim 0.64$ \AA. The two structures hence are gradually interpolated, moving the atoms along the direction defined by $\Vec{e}$. For each step, a self-consistent calculation with 500 eV for the kinetic energy cutoff and a $\Gamma$-centered 12$\times$12$\times$12 $k$-grid has been performed. The smearing here is constant and equal to 0.005 eV. The electronic calculations are considered converged below a threshold of $10^{-8}$ eV. This process has been studied while keeping the volume constant. Also here the spin-orbit coupling has not been included. To compute the frequency of the phonon mode we started from a quadratic fit around the minimum of the Born-Oppenheimer potential, corresponding to the CDW phase; the resulting fit has equation $y = 0.379 x^2$. From here it is possible to obtain the "spring" constant $k$, as $k = 2$$\times$0.379202 eV/\AA$^2$. The effective mass $m^*$ is computed using the normalized displacement vector $\hat{n} = \Vec{e}/\norm{\Vec{e}}$ and the mass tensor $\mathbf{M} =$ diag($m_1$, $m_1$, $m_1$, \dots, $m_{117}$, $m_{117}$, $m_{117}$) via $m^* = \Vec{e}\cdot(\mathbf{M} \cdot \Vec{e})$; we obtain $m^* = 91.02$ u. Note that this mass value is intermediate among the Sn and Sc ones, i.e. the atoms which mostly participate to the CDW transition. Finally the frequency $\nu$ can be computed as $\nu = \omega/(2\pi)$, with $\omega = \sqrt{k/m^*}$, giving us $\nu = 1.42$ THz.
When considering the unfolding of the CDW supercell, the Kohn–Sham wave functions are expanded in plane waves up to a 400 eV energy cutoff, with a $k$-mesh resolution for the self-consistent electronic structure calculations of 0.02 reciprocal Angstroms. For the non-self-consistent calculations, the $k$-mesh resolution corresponds to 0.01 reciprocal Angstroms. In this case, the spin-orbit coupling has been considered and included self-consistently.
The initial CDW supercell corresponds in every case to the experimental one \cite{Arachchige_2022}. Band structures have been visualized using the VASPKIT postprocessing tool \cite{vaspkit}. VESTA \cite{vesta} has been used to visualize the crystal structures.\\
\noindent{\bf Acknowledgements}
The authors acknowledge Andrea Cavalleri for the fruitful discussions on this work. The research leading to these results has received funding from the European Union's Horizon 2020 research and innovation program under the Marie Sk{\l}odowska-Curie Grant Agreement No. 897276. We gratefully acknowledge the Gauss Centre for Supercomputing e.V. (https://www.gauss-centre.eu) for funding this project by providing computing time on the GCS Supercomputer SuperMUC-NG at Leibniz Supercomputing Centre (https://www.lrz.de). We are grateful for funding support from the Deutsche Forschungsgemeinschaft (DFG, German Research Foundation) under Germany's Excellence Strategy through the W\"urzburg-Dresden Cluster of Excellence on Complexity and Topology in Quantum Matter ct.qmat (EXC 2147, Project ID 390858490) as well as through the Collaborative Research Center SFB 1170 ToCoTronics (Project ID 258499086) and the Hamburg Cluster of Excellence ``CUI: Advanced Imaging of Matter'' (EXC~2056, Project No.~390715994). This work has been performed in the framework of the Nanoscience Foundry and Fine Analysis (NFFA-MUR Italy Progetti Internazionali) facility. The Flatiron Institute is a division of the Simons Foundation. S.D.W. and G. Po. acknowledge support via the UC Santa Barbara NSF Quantum Foundry funded via the Q-AMASE-i program under award DMR-1906325. F.M. greatly acknowledges the SoE action of pnrr, number SOE\_0000068. I.Z. acknowledges the support from U.S. Department of Energy (DOE) Early Career Award DE-SC0020130.\\
\noindent{\bf Author contributions}
F.M., D.D.S and F.C. conceived and designed the project. G.Po. and S.W. grew the crystals. F.M., C.B., J.F., I.V., and P.C. carried out the ARPES measurements, while F.C., M.T., D.P., and W. B. obtained the pump and probe results. A.C. and S.E. performed the numerical calculations and theoretical analysis supervised by D.D.S, G.S., T.W. and R.T. All the authors participated in the discussion and crucially contributed to understanding and the writing of the manuscript.\\
\medskip
\noindent{\bf Additional information}
\medskip
\noindent{\bf Extended data} is available for this paper at https://xxxx\\
\noindent{\bf Supplementary information} The online version contains supplementary material available at https://xxx\\
\noindent{\bf Correspondence and requests for materials} should be addressed to Federico Mazzola, Domenico Di Sante and Federico Cilento.\\
\noindent{\bf Data availability} The data that support the findings of this study are available from the corresponding authors upon reasonable request.\\
\noindent{\bf Competing financial interests}\\
The authors declare no competing financial interests.

\bibliographystyle{naturemag}
\bibliography{biblio.bib}

\begin{thebibliography}{10}
\expandafter\ifx\csname url\endcsname\relax
  \def\url#1{\texttt{#1}}\fi
\expandafter\ifx\csname urlprefix\endcsname\relax\def\urlprefix{URL }\fi
\providecommand{\bibinfo}[2]{#2}
\providecommand{\eprint}[2][]{\url{#2}}

\bibitem{Neupert_2022}
\bibinfo{author}{Neupert, T.}, \bibinfo{author}{Denner, M.~M.},
  \bibinfo{author}{Yin, J.-X.}, \bibinfo{author}{Thomale, R.} \&
  \bibinfo{author}{Hasan, M.~Z.}
\newblock \bibinfo{title}{Charge order and superconductivity in kagome
  materials}.
\newblock \emph{\bibinfo{journal}{Nature Physics}}
  \textbf{\bibinfo{volume}{18}}, \bibinfo{pages}{137--143}
  (\bibinfo{year}{2022}).
\newblock \urlprefix\url{https://www.nature.com/articles/s41567-021-01404-y}.

\bibitem{Arachchige_2022}
\bibinfo{author}{Arachchige, H. W.~S.} \emph{et~al.}
\newblock \bibinfo{title}{Charge density wave in kagome lattice intermetallic
  {S}c{V}$_6${S}n$_6$}.
\newblock \emph{\bibinfo{journal}{Phys. Rev. Lett.}}
  \textbf{\bibinfo{volume}{129}}, \bibinfo{pages}{216402}
  (\bibinfo{year}{2022}).
\newblock
  \urlprefix\url{https://link.aps.org/doi/10.1103/PhysRevLett.129.216402}.

\bibitem{Wang_2021a}
\bibinfo{author}{Wang, Z.~X.} \emph{et~al.}
\newblock \bibinfo{title}{Unconventional charge density wave and photoinduced
  lattice symmetry change in the kagome metal {C}s{V}$_3${S}b$_5$ probed by
  time-resolved spectroscopy}.
\newblock \emph{\bibinfo{journal}{Phys. Rev. B}}
  \textbf{\bibinfo{volume}{104}}, \bibinfo{pages}{165110}
  (\bibinfo{year}{2021}).
\newblock \urlprefix\url{https://link.aps.org/doi/10.1103/PhysRevB.104.165110}.

\bibitem{Heumen_2021}
\bibinfo{author}{van Heumen, E.}
\newblock \bibinfo{title}{Kagome lattices with chiral charge density}.
\newblock \emph{\bibinfo{journal}{Nature Materials}}
  \textbf{\bibinfo{volume}{20}}, \bibinfo{pages}{1308--1309}
  (\bibinfo{year}{2021}).
\newblock \urlprefix\url{https://www.nature.com/articles/s41563-021-01095-z}.

\bibitem{Luo_2022}
\bibinfo{author}{Luo, H.} \emph{et~al.}
\newblock \bibinfo{title}{Electronic nature of charge density wave and
  electron-phonon coupling in kagome superconductor {K}{V}$_3${S}b$_5$}.
\newblock \emph{\bibinfo{journal}{Nature Communications}}
  \textbf{\bibinfo{volume}{13}}, \bibinfo{pages}{273} (\bibinfo{year}{2022}).
\newblock \urlprefix\url{https://doi.org/10.1038/s41467-021-27946-6}.

\bibitem{Kang_2020}
\bibinfo{author}{Kang, M.} \emph{et~al.}
\newblock \bibinfo{title}{Topological flat bands in frustrated kagome lattice
  {C}o{S}n}.
\newblock \emph{\bibinfo{journal}{Nature Communications}}
  \textbf{\bibinfo{volume}{11}}, \bibinfo{pages}{4004} (\bibinfo{year}{2020}).
\newblock \urlprefix\url{https://www.nature.com/articles/s41467-020-17465-1}.

\bibitem{Kang_2020b}
\bibinfo{author}{Kang, M.} \emph{et~al.}
\newblock \bibinfo{title}{Dirac fermions and flat bands in the ideal kagome
  metal {F}e{S}n}.
\newblock \emph{\bibinfo{journal}{Nature Materials}}
  \textbf{\bibinfo{volume}{19}}, \bibinfo{pages}{163--169}
  (\bibinfo{year}{2020}).
\newblock \urlprefix\url{https://www.nature.com/articles/s41563-019-0531-0}.

\bibitem{Lin_2018}
\bibinfo{author}{Lin, Z.} \emph{et~al.}
\newblock \bibinfo{title}{Flatbands and emergent ferromagnetic ordering in
  {F}e$_3${S}n$_2$ kagome lattices}.
\newblock \emph{\bibinfo{journal}{Phys. Rev. Lett.}}
  \textbf{\bibinfo{volume}{121}}, \bibinfo{pages}{096401}
  (\bibinfo{year}{2018}).
\newblock
  \urlprefix\url{https://journals.aps.org/prl/abstract/10.1103/PhysRevLett.121.096401}.

\bibitem{Ye_2019}
\bibinfo{author}{Ye, L.} \emph{et~al.}
\newblock \bibinfo{title}{de {H}aas-van {A}lphen effect of correlated {D}irac
  states in kagome metal {F}e$_3${S}n$_2$}.
\newblock \emph{\bibinfo{journal}{Nature Communications}}
  \textbf{\bibinfo{volume}{10}}, \bibinfo{pages}{4870} (\bibinfo{year}{2019}).
\newblock \urlprefix\url{https://www.nature.com/articles/s41467-019-12822-1}.

\bibitem{Pokharel_2021}
\bibinfo{author}{Pokharel, G.} \emph{et~al.}
\newblock \bibinfo{title}{Electronic properties of the topological kagome
  metals {Y}{V}$_6${S}n$_6$ and {Gd}{V}$_6${S}n$_6$}.
\newblock \emph{\bibinfo{journal}{Phys. Rev. B}}
  \textbf{\bibinfo{volume}{104}}, \bibinfo{pages}{235139}
  (\bibinfo{year}{2021}).
\newblock
  \urlprefix\url{https://journals.aps.org/prb/abstract/10.1103/PhysRevB.104.235139}.

\bibitem{Hu_2022}
\bibinfo{author}{Hu, Y.} \emph{et~al.}
\newblock \bibinfo{title}{Tunable topological {D}irac surface states and van
  {H}ove singularities in kagome metal {Gd}{V}$_6${S}n$_6$}.
\newblock \emph{\bibinfo{journal}{Science Advances}}
  \textbf{\bibinfo{volume}{8}}, \bibinfo{pages}{eadd2024}
  (\bibinfo{year}{2022}).
\newblock \urlprefix\url{https://www.science.org/doi/10.1126/sciadv.add2024}.

\bibitem{Kang_2023}
\bibinfo{author}{Kang, M.} \emph{et~al.}
\newblock \bibinfo{title}{Charge order landscape and competition with
  superconductivity in kagome metals}.
\newblock \emph{\bibinfo{journal}{Nature Materials}}
  \textbf{\bibinfo{volume}{22}}, \bibinfo{pages}{186--193}
  (\bibinfo{year}{2023}).
\newblock \urlprefix\url{https://doi.org/10.1038/s41563-022-01375-2}.

\bibitem{Teng_2022}
\bibinfo{author}{Teng, X.} \emph{et~al.}
\newblock \bibinfo{title}{Discovery of charge density wave in a kagome lattice
  antiferromagnet}.
\newblock \emph{\bibinfo{journal}{Nature}} \textbf{\bibinfo{volume}{609}},
  \bibinfo{pages}{490--495} (\bibinfo{year}{2022}).
\newblock \urlprefix\url{https://doi.org/10.1038/s41586-022-05034-z}.

\bibitem{Peng_2021}
\bibinfo{author}{Peng, S.} \emph{et~al.}
\newblock \bibinfo{title}{Realizing kagome band structure in two-dimensional
  kagome surface states of {$R$}{V}$_6${S}n$_6$({$R$}= {Gd}, {H}o)}.
\newblock \emph{\bibinfo{journal}{Phys. Rev. Lett.}}
  \textbf{\bibinfo{volume}{127}}, \bibinfo{pages}{266401}
  (\bibinfo{year}{2021}).
\newblock
  \urlprefix\url{https://journals.aps.org/prl/abstract/10.1103/PhysRevLett.127.266401}.

\bibitem{Ishikawa_2021}
\bibinfo{author}{Ishikawa, H.}, \bibinfo{author}{Yajima, T.},
  \bibinfo{author}{Kawamura, M.}, \bibinfo{author}{Mitamura, H.} \&
  \bibinfo{author}{Kindo, K.}
\newblock \bibinfo{title}{{G}d{V}$_6${S}n$_6$: A {M}ulti-carrier {M}etal with
  {N}on-magnetic 3$d$-electron {K}agome {B}ands and 4$f$-electron {M}agnetism}.
\newblock \emph{\bibinfo{journal}{Journal of the Physical Society of Japan}}
  \textbf{\bibinfo{volume}{90}}, \bibinfo{pages}{124704}
  (\bibinfo{year}{2021}).
\newblock \urlprefix\url{https://journals.jps.jp/doi/10.7566/JPSJ.90.124704}.

\bibitem{Rosenberg_2022}
\bibinfo{author}{Rosenberg, E.} \emph{et~al.}
\newblock \bibinfo{title}{Uniaxial ferromagnetism in the kagome metal
  {T}b{V}$_6${S}n$_6$}.
\newblock \emph{\bibinfo{journal}{Phys. Rev. B}}
  \textbf{\bibinfo{volume}{106}}, \bibinfo{pages}{115139}
  (\bibinfo{year}{2022}).
\newblock \urlprefix\url{https://link.aps.org/doi/10.1103/PhysRevB.106.115139}.

\bibitem{Lee_2022}
\bibinfo{author}{Lee, J.} \& \bibinfo{author}{Mun, E.}
\newblock \bibinfo{title}{Anisotropic magnetic property of single crystals
  {R}{V}$_6${S}n$_6$ $({R}=\mathrm{Y},
  \mathrm{Gd}\text{\ensuremath{-}}\mathrm{Tm}, \mathrm{Lu})$}.
\newblock \emph{\bibinfo{journal}{Phys. Rev. Mater.}}
  \textbf{\bibinfo{volume}{6}}, \bibinfo{pages}{083401} (\bibinfo{year}{2022}).
\newblock
  \urlprefix\url{https://link.aps.org/doi/10.1103/PhysRevMaterials.6.083401}.

\bibitem{Chen_2021}
\bibinfo{author}{Chen, D.} \emph{et~al.}
\newblock \bibinfo{title}{Large anomalous hall effect in the kagome ferromagnet
  {L}i{M}n$_6${S}n$_6$}.
\newblock \emph{\bibinfo{journal}{Phys. Rev. B}}
  \textbf{\bibinfo{volume}{103}}, \bibinfo{pages}{144410}
  (\bibinfo{year}{2021}).
\newblock
  \urlprefix\url{https://journals.aps.org/prb/abstract/10.1103/PhysRevB.103.144410}.

\bibitem{Guo2009}
\bibinfo{author}{Guo, H.-M.} \& \bibinfo{author}{Franz, M.}
\newblock \bibinfo{title}{Topological insulator on the kagome lattice}.
\newblock \emph{\bibinfo{journal}{Phys. Rev. B}} \textbf{\bibinfo{volume}{80}},
  \bibinfo{pages}{113102} (\bibinfo{year}{2009}).
\newblock
  \urlprefix\url{https://journals.aps.org/prb/abstract/10.1103/PhysRevB.80.113102}.

\bibitem{Mazin2014}
\bibinfo{author}{Mazin, I.~I.} \emph{et~al.}
\newblock \bibinfo{title}{Theoretical prediction of a strongly correlated
  {D}irac metal}.
\newblock \emph{\bibinfo{journal}{Nature Communications}}
  \textbf{\bibinfo{volume}{5}}, \bibinfo{pages}{4261} (\bibinfo{year}{2014}).
\newblock \urlprefix\url{https://www.nature.com/articles/ncomms5261}.

\bibitem{Kiesel2013}
\bibinfo{author}{Kiesel, M.~L.}, \bibinfo{author}{Platt, C.} \&
  \bibinfo{author}{Thomale, R.}
\newblock \bibinfo{title}{Unconventional {F}ermi {S}urface {I}nstabilities in
  the {K}agome {H}ubbard {M}odel}.
\newblock \emph{\bibinfo{journal}{Phys. Rev. Lett.}}
  \textbf{\bibinfo{volume}{110}}, \bibinfo{pages}{126405}
  (\bibinfo{year}{2013}).
\newblock
  \urlprefix\url{https://journals.aps.org/prl/abstract/10.1103/PhysRevLett.110.126405}.

\bibitem{DiSante2020}
\bibinfo{author}{Di~Sante, D.} \emph{et~al.}
\newblock \bibinfo{title}{Turbulent hydrodynamics in strongly correlated
  {K}agome metals}.
\newblock \emph{\bibinfo{journal}{Nature Communications}}
  \textbf{\bibinfo{volume}{11}}, \bibinfo{pages}{3997} (\bibinfo{year}{2020}).
\newblock \urlprefix\url{https://www.nature.com/articles/s41467-020-17663-x}.

\bibitem{Guo_2022}
\bibinfo{author}{Guo, C.} \emph{et~al.}
\newblock \bibinfo{title}{Switchable chiral transport in charge-ordered kagome
  metal {C}s{V}$_3${S}b$_5$}.
\newblock \emph{\bibinfo{journal}{Nature}} \textbf{\bibinfo{volume}{611}},
  \bibinfo{pages}{461--466} (\bibinfo{year}{2022}).
\newblock \urlprefix\url{https://doi.org/10.1038/s41586-022-05127-9}.

\bibitem{Nishimoto_2010}
\bibinfo{author}{Nishimoto, S.}, \bibinfo{author}{Nakamura, M.},
  \bibinfo{author}{O'Brien, A.} \& \bibinfo{author}{Fulde, P.}
\newblock \bibinfo{title}{Metal-insulator transition of fermions on a kagome
  lattice at 1/3 filling}.
\newblock \emph{\bibinfo{journal}{Phys. Rev. Lett.}}
  \textbf{\bibinfo{volume}{104}}, \bibinfo{pages}{196401}
  (\bibinfo{year}{2010}).
\newblock
  \urlprefix\url{https://link.aps.org/doi/10.1103/PhysRevLett.104.196401}.

\bibitem{Kiesel_2013}
\bibinfo{author}{Kiesel, M.~L.}, \bibinfo{author}{Platt, C.} \&
  \bibinfo{author}{Thomale, R.}
\newblock \bibinfo{title}{Unconventional fermi surface instabilities in the
  kagome hubbard model}.
\newblock \emph{\bibinfo{journal}{Phys. Rev. Lett.}}
  \textbf{\bibinfo{volume}{110}}, \bibinfo{pages}{126405}
  (\bibinfo{year}{2013}).
\newblock
  \urlprefix\url{https://link.aps.org/doi/10.1103/PhysRevLett.110.126405}.

\bibitem{Denner_2021}
\bibinfo{author}{Denner, M.~M.}, \bibinfo{author}{Thomale, R.} \&
  \bibinfo{author}{Neupert, T.}
\newblock \bibinfo{title}{Analysis of {C}harge {O}rder in the {K}agome {M}etal
  {$A$}{V}$_3${S}b$_5$ (${A}=\mathrm{K},\mathrm{Rb},\mathrm{Cs}$)}.
\newblock \emph{\bibinfo{journal}{Phys. Rev. Lett.}}
  \textbf{\bibinfo{volume}{127}}, \bibinfo{pages}{217601}
  (\bibinfo{year}{2021}).
\newblock
  \urlprefix\url{https://link.aps.org/doi/10.1103/PhysRevLett.127.217601}.

\bibitem{Feng_2021}
\bibinfo{author}{Feng, X.}, \bibinfo{author}{Zhang, Y.},
  \bibinfo{author}{Jiang, K.} \& \bibinfo{author}{Hu, J.}
\newblock \bibinfo{title}{Low-energy effective theory and symmetry
  classification of flux phases on the kagome lattice}.
\newblock \emph{\bibinfo{journal}{Phys. Rev. B}}
  \textbf{\bibinfo{volume}{104}}, \bibinfo{pages}{165136}
  (\bibinfo{year}{2021}).
\newblock \urlprefix\url{https://link.aps.org/doi/10.1103/PhysRevB.104.165136}.

\bibitem{Park_2021}
\bibinfo{author}{Park, T.}, \bibinfo{author}{Ye, M.} \&
  \bibinfo{author}{Balents, L.}
\newblock \bibinfo{title}{Electronic instabilities of kagome metals: {S}addle
  points and {L}andau theory}.
\newblock \emph{\bibinfo{journal}{Phys. Rev. B}}
  \textbf{\bibinfo{volume}{104}}, \bibinfo{pages}{035142}
  (\bibinfo{year}{2021}).
\newblock \urlprefix\url{https://link.aps.org/doi/10.1103/PhysRevB.104.035142}.

\bibitem{Jiang_2021}
\bibinfo{author}{Jiang, Y.-X.} \emph{et~al.}
\newblock \bibinfo{title}{Unconventional chiral charge order in kagome
  superconductor {K}{V}$_3${S}b$_5$}.
\newblock \emph{\bibinfo{journal}{Nature Materials}}
  \textbf{\bibinfo{volume}{20}}, \bibinfo{pages}{1353--1357}
  (\bibinfo{year}{2021}).
\newblock \urlprefix\url{https://doi.org/10.1038/s41563-021-01034-y}.

\bibitem{Liang_2021}
\bibinfo{author}{Liang, Z.} \emph{et~al.}
\newblock \bibinfo{title}{Three-dimensional charge density wave and
  surface-dependent vortex-core states in a kagome superconductor
  {C}s{V}$_3${S}b$_5$}.
\newblock \emph{\bibinfo{journal}{Phys. Rev. X}} \textbf{\bibinfo{volume}{11}},
  \bibinfo{pages}{031026} (\bibinfo{year}{2021}).
\newblock \urlprefix\url{https://link.aps.org/doi/10.1103/PhysRevX.11.031026}.

\bibitem{Wu2021}
\bibinfo{author}{Wu, X.} \emph{et~al.}
\newblock \bibinfo{title}{Nature of {U}nconventional {P}airing in the {K}agome
  {S}uperconductors {$A$}{V}$_3${S}b$_5$ (${A}= \mathrm{K}, \mathrm{Rb},
  \mathrm{Cs}$)}.
\newblock \emph{\bibinfo{journal}{Phys. Rev. Lett.}}
  \textbf{\bibinfo{volume}{127}}, \bibinfo{pages}{177001}
  (\bibinfo{year}{2021}).
\newblock
  \urlprefix\url{https://journals.aps.org/prl/abstract/10.1103/PhysRevLett.127.177001}.

\bibitem{Kang2022}
\bibinfo{author}{Kang, M.} \emph{et~al.}
\newblock \bibinfo{title}{Twofold van {H}ove singularity and origin of charge
  order in topological kagome superconductor {C}s{V}$_3${S}b$_5$}.
\newblock \emph{\bibinfo{journal}{Nature Physics}}
  \textbf{\bibinfo{volume}{18}}, \bibinfo{pages}{301--308}
  (\bibinfo{year}{2022}).
\newblock \urlprefix\url{https://www.nature.com/articles/s41567-021-01451-5}.

\bibitem{Ortiz2020}
\bibinfo{author}{Ortiz, B.~R.} \emph{et~al.}
\newblock \bibinfo{title}{{C}s{V}$_3${S}b$_5$: A {Z}$_{2}$ topological kagome
  metal with a superconducting ground state}.
\newblock \emph{\bibinfo{journal}{Phys. Rev. Lett.}}
  \textbf{\bibinfo{volume}{125}}, \bibinfo{pages}{247002}
  (\bibinfo{year}{2020}).
\newblock
  \urlprefix\url{https://journals.aps.org/prl/abstract/10.1103/PhysRevLett.125.247002}.

\bibitem{Yu_2021}
\bibinfo{author}{Yu, F.~H.} \emph{et~al.}
\newblock \bibinfo{title}{Unusual competition of superconductivity and
  charge-density-wave state in a compressed topological kagome metal}.
\newblock \emph{\bibinfo{journal}{Nature Communications}}
  \textbf{\bibinfo{volume}{12}}, \bibinfo{pages}{3645} (\bibinfo{year}{2021}).
\newblock \urlprefix\url{https://doi.org/10.1038/s41467-021-23928-w}.

\bibitem{Du_2021}
\bibinfo{author}{Du, F.} \emph{et~al.}
\newblock \bibinfo{title}{Pressure-induced double superconducting domes and
  charge instability in the kagome metal {K}{V}$_3${S}b$_5$}.
\newblock \emph{\bibinfo{journal}{Phys. Rev. B}}
  \textbf{\bibinfo{volume}{103}}, \bibinfo{pages}{L220504}
  (\bibinfo{year}{2021}).
\newblock
  \urlprefix\url{https://link.aps.org/doi/10.1103/PhysRevB.103.L220504}.

\bibitem{Oey_2022}
\bibinfo{author}{Oey, Y.~M.} \emph{et~al.}
\newblock \bibinfo{title}{Fermi level tuning and double-dome superconductivity
  in the kagome metal {C}s{V}$_3${S}b$_{5-x}${S}n$_{x}$}.
\newblock \emph{\bibinfo{journal}{Phys. Rev. Mater.}}
  \textbf{\bibinfo{volume}{6}}, \bibinfo{pages}{L041801}
  (\bibinfo{year}{2022}).
\newblock
  \urlprefix\url{https://link.aps.org/doi/10.1103/PhysRevMaterials.6.L041801}.

\bibitem{Li_2021a}
\bibinfo{author}{Li, H.} \emph{et~al.}
\newblock \bibinfo{title}{Observation of {U}nconventional {C}harge {D}ensity
  {W}ave without {A}coustic {P}honon {A}nomaly in {K}agome {S}uperconductors
  {$A$}{V}$_3${S}b$_5$ (${A}$ = {Rb}, {C}s)}.
\newblock \emph{\bibinfo{journal}{Phys. Rev. X}} \textbf{\bibinfo{volume}{11}},
  \bibinfo{pages}{031050} (\bibinfo{year}{2021}).
\newblock \urlprefix\url{https://link.aps.org/doi/10.1103/PhysRevX.11.031050}.

\bibitem{Wang_2021b}
\bibinfo{author}{Wang, Z.} \emph{et~al.}
\newblock \bibinfo{title}{Electronic nature of chiral charge order in the
  kagome superconductor {C}s{V}$_3${S}b$_5$}.
\newblock \emph{\bibinfo{journal}{Phys. Rev. B}}
  \textbf{\bibinfo{volume}{104}}, \bibinfo{pages}{075148}
  (\bibinfo{year}{2021}).
\newblock \urlprefix\url{https://link.aps.org/doi/10.1103/PhysRevB.104.075148}.

\bibitem{Zhao_2021}
\bibinfo{author}{Zhao, H.} \emph{et~al.}
\newblock \bibinfo{title}{Cascade of correlated electron states in the kagome
  superconductor {C}s{V}$_3${S}b$_5$}.
\newblock \emph{\bibinfo{journal}{Nature}} \textbf{\bibinfo{volume}{599}},
  \bibinfo{pages}{216--221} (\bibinfo{year}{2021}).
\newblock \urlprefix\url{https://doi.org/10.1038/s41586-021-03946-w}.

\bibitem{Ortiz_2021}
\bibinfo{author}{Ortiz, B.~R.} \emph{et~al.}
\newblock \bibinfo{title}{Fermi surface mapping and the nature of
  charge-density-wave order in the kagome superconductor {C}s{V}$_3${S}b$_5$}.
\newblock \emph{\bibinfo{journal}{Phys. Rev. X}} \textbf{\bibinfo{volume}{11}},
  \bibinfo{pages}{041030} (\bibinfo{year}{2021}).
\newblock \urlprefix\url{https://link.aps.org/doi/10.1103/PhysRevX.11.041030}.

\bibitem{Nie_2022}
\bibinfo{author}{Nie, L.} \emph{et~al.}
\newblock \bibinfo{title}{Charge-density-wave-driven electronic nematicity in a
  kagome superconductor}.
\newblock \emph{\bibinfo{journal}{Nature}} \textbf{\bibinfo{volume}{604}},
  \bibinfo{pages}{59--64} (\bibinfo{year}{2022}).
\newblock \urlprefix\url{https://doi.org/10.1038/s41586-022-04493-8}.

\bibitem{Zhang_2022}
\bibinfo{author}{Zhang, X.} \emph{et~al.}
\newblock \bibinfo{title}{Destabilization of the charge density wave and the
  absence of superconductivity in {S}c{V}$_6${S}n$_6$ under high pressures up
  to 11 {G}{P}a}.
\newblock \emph{\bibinfo{journal}{Materials}} \textbf{\bibinfo{volume}{15}}
  (\bibinfo{year}{2022}).
\newblock \urlprefix\url{https://www.mdpi.com/1996-1944/15/20/7372}.

\bibitem{Yin_2022}
\bibinfo{author}{Yin, J.-X.}, \bibinfo{author}{Lian, B.} \&
  \bibinfo{author}{Hasan, M.~Z.}
\newblock \bibinfo{title}{Topological kagome magnets and superconductors}.
\newblock \emph{\bibinfo{journal}{Nature}} \textbf{\bibinfo{volume}{612}},
  \bibinfo{pages}{647--657} (\bibinfo{year}{2022}).
\newblock \urlprefix\url{https://doi.org/10.1038/s41586-022-05516-0}.

\bibitem{Sun_2022}
\bibinfo{author}{Sun, Z.} \emph{et~al.}
\newblock \bibinfo{title}{Observation of topological flat bands in the kagome
  semiconductor {N}b$_3${C}l$_8$}.
\newblock \emph{\bibinfo{journal}{Nano Letters}} \textbf{\bibinfo{volume}{22}},
  \bibinfo{pages}{4596--4602} (\bibinfo{year}{2022}).
\newblock
  \urlprefix\url{https://pubs.acs.org/doi/10.1021/acs.nanolett.2c00778}.

\bibitem{Tanaka_2020a}
\bibinfo{author}{Tanaka, M.} \emph{et~al.}
\newblock \bibinfo{title}{Topological kagome magnet {C}o$_3${S}n$_2${S}$_2$
  thin flakes with high electron mobility and large anomalous hall effect}.
\newblock \emph{\bibinfo{journal}{Nano Letters}} \textbf{\bibinfo{volume}{20}},
  \bibinfo{pages}{7476--7481} (\bibinfo{year}{2020}).
\newblock \urlprefix\url{https://doi.org/10.1021/acs.nanolett.0c02962}.

\bibitem{NagaosaPRB2019}
\bibinfo{author}{Bolens, A.} \& \bibinfo{author}{Nagaosa, N.}
\newblock \bibinfo{title}{Topological states on the breathing kagome lattice}.
\newblock \emph{\bibinfo{journal}{Phys. Rev. B}} \textbf{\bibinfo{volume}{99}},
  \bibinfo{pages}{165141} (\bibinfo{year}{2019}).
\newblock \urlprefix\url{https://link.aps.org/doi/10.1103/PhysRevB.99.165141}.

\bibitem{Whangbo_1991}
\bibinfo{author}{Whangbo, M.-H.}, \bibinfo{author}{Canadell, E.},
  \bibinfo{author}{Foury, P.} \& \bibinfo{author}{Pouget, J.-P.}
\newblock \bibinfo{title}{Hidden fermi surface nesting and charge density wave
  instability in low-dimensional metals}.
\newblock \emph{\bibinfo{journal}{Science}} \textbf{\bibinfo{volume}{252}},
  \bibinfo{pages}{96--98} (\bibinfo{year}{1991}).
\newblock
  \urlprefix\url{https://www.science.org/doi/abs/10.1126/science.252.5002.96}.

\bibitem{Carpinelli_1996}
\bibinfo{author}{Carpinelli, J.~M.}, \bibinfo{author}{Weitering, H.~H.},
  \bibinfo{author}{Plummer, E.~W.} \& \bibinfo{author}{Stumpf, R.}
\newblock \bibinfo{title}{Direct observation of a surface charge density wave}.
\newblock \emph{\bibinfo{journal}{Nature}} \textbf{\bibinfo{volume}{381}},
  \bibinfo{pages}{398--400} (\bibinfo{year}{1996}).
\newblock \urlprefix\url{https://doi.org/10.1038/381398a0}.

\bibitem{Tam_2022}
\bibinfo{author}{Tam, C.~C.} \emph{et~al.}
\newblock \bibinfo{title}{Charge density waves and fermi surface reconstruction
  in the clean overdoped cuprate superconductor
  {T}l$_2${B}a$_2${C}u{O}$_{6+\delta}$}.
\newblock \emph{\bibinfo{journal}{Nature Communications}}
  \textbf{\bibinfo{volume}{13}}, \bibinfo{pages}{570} (\bibinfo{year}{2022}).
\newblock \urlprefix\url{https://doi.org/10.1038/s41467-022-28124-y}.

\bibitem{Bosak_2021}
\bibinfo{author}{Bosak, A.} \emph{et~al.}
\newblock \bibinfo{title}{Evidence for nesting-driven charge density wave
  instabilities in the quasi-two-dimensional material {L}a{A}g{S}b$_2$}.
\newblock \emph{\bibinfo{journal}{Phys. Rev. Res.}}
  \textbf{\bibinfo{volume}{3}}, \bibinfo{pages}{033020} (\bibinfo{year}{2021}).
\newblock
  \urlprefix\url{https://link.aps.org/doi/10.1103/PhysRevResearch.3.033020}.

\bibitem{Sunko_2020}
\bibinfo{author}{Sunko, V.} \emph{et~al.}
\newblock \bibinfo{title}{Probing spin correlations using angle-resolved
  photoemission in a coupled metallic/mott insulator system}.
\newblock \emph{\bibinfo{journal}{Science Advances}}
  \textbf{\bibinfo{volume}{6}}, \bibinfo{pages}{eaaz0611}
  (\bibinfo{year}{2020}).
\newblock
  \urlprefix\url{https://www.science.org/doi/abs/10.1126/sciadv.aaz0611}.

\bibitem{Li_2014}
\bibinfo{author}{Li, C.~H.} \emph{et~al.}
\newblock \bibinfo{title}{Electrical detection of charge-current-induced spin
  polarization due to spin-momentum locking in {B}i$_2${S}e$_3$}.
\newblock \emph{\bibinfo{journal}{Nature Nanotechnology}}
  \textbf{\bibinfo{volume}{9}}, \bibinfo{pages}{218--224}
  (\bibinfo{year}{2014}).
\newblock \urlprefix\url{https://doi.org/10.1038/nnano.2014.16}.

\bibitem{Li_2018}
\bibinfo{author}{Li, P.} \emph{et~al.}
\newblock \bibinfo{title}{Spin-momentum locking and spin-orbit torques in
  magnetic nano-heterojunctions composed of weyl semimetal {W}{T}e$_2$}.
\newblock \emph{\bibinfo{journal}{Nature Communications}}
  \textbf{\bibinfo{volume}{9}}, \bibinfo{pages}{3990} (\bibinfo{year}{2018}).
\newblock \urlprefix\url{https://doi.org/10.1038/s41467-018-06518-1}.

\bibitem{Jiang_2016}
\bibinfo{author}{Jiang, Z.} \emph{et~al.}
\newblock \bibinfo{title}{Enhanced spin seebeck effect signal due to
  spin-momentum locked topological surface states}.
\newblock \emph{\bibinfo{journal}{Nature Communications}}
  \textbf{\bibinfo{volume}{7}}, \bibinfo{pages}{11458} (\bibinfo{year}{2016}).
\newblock \urlprefix\url{https://doi.org/10.1038/ncomms11458}.

\bibitem{Luo_2017}
\bibinfo{author}{Luo, S.}, \bibinfo{author}{He, L.} \& \bibinfo{author}{Li, M.}
\newblock \bibinfo{title}{Spin-momentum locked interaction between guided
  photons and surface electrons in topological insulators}.
\newblock \emph{\bibinfo{journal}{Nature Communications}}
  \textbf{\bibinfo{volume}{8}}, \bibinfo{pages}{2141} (\bibinfo{year}{2017}).
\newblock \urlprefix\url{https://doi.org/10.1038/s41467-017-02264-y}.

\bibitem{Li_2014b}
\bibinfo{author}{Li, Q.} \emph{et~al.}
\newblock \bibinfo{title}{Bond competition and phase evolution on the
  {I}r{T}e$_2$ surface}.
\newblock \emph{\bibinfo{journal}{Nature Communications}}
  \textbf{\bibinfo{volume}{5}}, \bibinfo{pages}{5358} (\bibinfo{year}{2014}).
\newblock \urlprefix\url{https://doi.org/10.1038/ncomms6358}.

\bibitem{Fang_2013}
\bibinfo{author}{Fang, A.~F.}, \bibinfo{author}{Xu, G.}, \bibinfo{author}{Dong,
  T.}, \bibinfo{author}{Zheng, P.} \& \bibinfo{author}{Wang, N.~L.}
\newblock \bibinfo{title}{Structural phase transition in {I}r{T}e$_2$: A
  combined study of optical spectroscopy and band structure calculations}.
\newblock \emph{\bibinfo{journal}{Scientific Reports}}
  \textbf{\bibinfo{volume}{3}}, \bibinfo{pages}{1153} (\bibinfo{year}{2013}).
\newblock \urlprefix\url{https://doi.org/10.1038/srep01153}.

\bibitem{Tomeljak_2009}
\bibinfo{author}{Tomeljak, A.} \emph{et~al.}
\newblock \bibinfo{title}{Dynamics of photoinduced charge-density-wave to metal
  phase transition in {K}$_0.3${Mo}{O}$_3$}.
\newblock \emph{\bibinfo{journal}{Phys. Rev. Lett.}}
  \textbf{\bibinfo{volume}{102}}, \bibinfo{pages}{066404}
  (\bibinfo{year}{2009}).
\newblock
  \urlprefix\url{https://link.aps.org/doi/10.1103/PhysRevLett.102.066404}.

\bibitem{Yusupov_2008}
\bibinfo{author}{Yusupov, R.~V.}, \bibinfo{author}{Mertelj, T.},
  \bibinfo{author}{Chu, J.-H.}, \bibinfo{author}{Fisher, I.~R.} \&
  \bibinfo{author}{Mihailovic, D.}
\newblock \bibinfo{title}{Single-particle and collective mode couplings
  associated with 1- and 2-directional electronic ordering in metallic
  {R}{T}e$_3$ ({R}={H}o, {D}y, {T}b}.
\newblock \emph{\bibinfo{journal}{Phys. Rev. Lett.}}
  \textbf{\bibinfo{volume}{101}}, \bibinfo{pages}{246402}
  (\bibinfo{year}{2008}).
\newblock
  \urlprefix\url{https://link.aps.org/doi/10.1103/PhysRevLett.101.246402}.

\bibitem{Pokharel2022}
\bibinfo{author}{Pokharel, A.~R.} \emph{et~al.}
\newblock \bibinfo{title}{Dynamics of collective modes in an unconventional
  charge density wave system {B}a{N}i$_2${A}s$_2$}.
\newblock \emph{\bibinfo{journal}{Communications Physics}}
  \textbf{\bibinfo{volume}{5}}, \bibinfo{pages}{141} (\bibinfo{year}{2022}).
\newblock \urlprefix\url{https://doi.org/10.1038/s42005-022-00919-x}.

\bibitem{Vorobeva_2011}
\bibinfo{author}{M\"ohr-Vorobeva, E.} \emph{et~al.}
\newblock \bibinfo{title}{Nonthermal melting of a charge density wave in
  {T}i{S}e$_2$}.
\newblock \emph{\bibinfo{journal}{Phys. Rev. Lett.}}
  \textbf{\bibinfo{volume}{107}}, \bibinfo{pages}{036403}
  (\bibinfo{year}{2011}).
\newblock
  \urlprefix\url{https://link.aps.org/doi/10.1103/PhysRevLett.107.036403}.

\bibitem{Ravnik_2018}
\bibinfo{author}{Ravnik, J.}, \bibinfo{author}{Vaskivskyi, I.},
  \bibinfo{author}{Mertelj, T.} \& \bibinfo{author}{Mihailovic, D.}
\newblock \bibinfo{title}{Real-time observation of the coherent transition to a
  metastable emergent state in 1{T}-{T}a{S}$_2$}.
\newblock \emph{\bibinfo{journal}{Phys. Rev. B}} \textbf{\bibinfo{volume}{97}},
  \bibinfo{pages}{075304} (\bibinfo{year}{2018}).
\newblock \urlprefix\url{https://link.aps.org/doi/10.1103/PhysRevB.97.075304}.

\bibitem{Eichberger2010}
\bibinfo{author}{Eichberger, M.} \emph{et~al.}
\newblock \bibinfo{title}{Snapshots of cooperative atomic motions in the
  optical suppression of charge density waves}.
\newblock \emph{\bibinfo{journal}{Nature}} \textbf{\bibinfo{volume}{468}},
  \bibinfo{pages}{799--802} (\bibinfo{year}{2010}).
\newblock \urlprefix\url{https://doi.org/10.1038/nature09539}.

\bibitem{TOMELJAK2009_physB}
\bibinfo{author}{Tomeljak, A.} \emph{et~al.}
\newblock \bibinfo{title}{Femtosecond nonequilibrium dynamics in quasi-1d cdw
  systems {K}$_{0.3}${M}o{O}$_3$ and {R}{B}$_{0.3}${M}o{O}$_3$}.
\newblock \emph{\bibinfo{journal}{Physica B: Condensed Matter}}
  \textbf{\bibinfo{volume}{404}}, \bibinfo{pages}{548--551}
  (\bibinfo{year}{2009}).
\newblock
  \urlprefix\url{https://www.sciencedirect.com/science/article/pii/S092145260800584X}.

\bibitem{Huber_2014}
\bibinfo{author}{Huber, T.} \emph{et~al.}
\newblock \bibinfo{title}{Coherent structural dynamics of a prototypical
  charge-density-wave-to-metal transition}.
\newblock \emph{\bibinfo{journal}{Phys. Rev. Lett.}}
  \textbf{\bibinfo{volume}{113}}, \bibinfo{pages}{026401}
  (\bibinfo{year}{2014}).
\newblock
  \urlprefix\url{https://link.aps.org/doi/10.1103/PhysRevLett.113.026401}.

\bibitem{Storeck_2020}
\bibinfo{author}{Storeck, G.} \emph{et~al.}
\newblock \bibinfo{title}{Structural dynamics of incommensurate charge-density
  waves tracked by ultrafast low-energy electron diffraction}.
\newblock \emph{\bibinfo{journal}{Structural Dynamics}}
  \textbf{\bibinfo{volume}{7}}, \bibinfo{pages}{034304} (\bibinfo{year}{2020}).
\newblock \urlprefix\url{https://doi.org/10.1063/4.0000018}.
\newblock \eprint{https://doi.org/10.1063/4.0000018}.

\bibitem{Demsar_1999}
\bibinfo{author}{Demsar, J.}, \bibinfo{author}{Biljakovi\ifmmode~\acute{c}\else
  \'{c}\fi{}, K.} \& \bibinfo{author}{Mihailovic, D.}
\newblock \bibinfo{title}{Single particle and collective excitations in the
  one-dimensional charge density wave solid {K}$_{0.3}${M}o{O}$_3$ probed in
  real time by femtosecond spectroscopy}.
\newblock \emph{\bibinfo{journal}{Phys. Rev. Lett.}}
  \textbf{\bibinfo{volume}{83}}, \bibinfo{pages}{800--803}
  (\bibinfo{year}{1999}).
\newblock \urlprefix\url{https://link.aps.org/doi/10.1103/PhysRevLett.83.800}.

\bibitem{Chen_2017}
\bibinfo{author}{Chen, R.~Y.}, \bibinfo{author}{Zhang, S.~J.},
  \bibinfo{author}{Zhang, M.~Y.}, \bibinfo{author}{Dong, T.} \&
  \bibinfo{author}{Wang, N.~L.}
\newblock \bibinfo{title}{Revealing extremely low energy amplitude modes in the
  charge-density-wave compound {L}a{A}g{S}b$_2$}.
\newblock \emph{\bibinfo{journal}{Phys. Rev. Lett.}}
  \textbf{\bibinfo{volume}{118}}, \bibinfo{pages}{107402}
  (\bibinfo{year}{2017}).
\newblock
  \urlprefix\url{https://link.aps.org/doi/10.1103/PhysRevLett.118.107402}.

\bibitem{Yusupov2010}
\bibinfo{author}{Yusupov, R.} \emph{et~al.}
\newblock \bibinfo{title}{Coherent dynamics of macroscopic electronic order
  through a symmetry breaking transition}.
\newblock \emph{\bibinfo{journal}{Nature Physics}}
  \textbf{\bibinfo{volume}{6}}, \bibinfo{pages}{681--684}
  (\bibinfo{year}{2010}).
\newblock \urlprefix\url{https://doi.org/10.1038/nphys1738}.

\bibitem{Yoshikawa2021}
\bibinfo{author}{Yoshikawa, N.} \emph{et~al.}
\newblock \bibinfo{title}{Ultrafast switching to an insulating-like metastable
  state by amplitudon excitation of a charge density wave}.
\newblock \emph{\bibinfo{journal}{Nature Physics}}
  \textbf{\bibinfo{volume}{17}}, \bibinfo{pages}{909--914}
  (\bibinfo{year}{2021}).
\newblock \urlprefix\url{https://doi.org/10.1038/s41567-021-01267-3}.

\bibitem{Schaefer2013}
\bibinfo{author}{Schaefer, H.} \emph{et~al.}
\newblock \bibinfo{title}{Dynamics of charge density wave order in the quasi
  one dimensional conductor ({T}a{S}e$_4$)$_2${I} probed by femtosecond optical
  spectroscopy}.
\newblock \emph{\bibinfo{journal}{The European Physical Journal Special
  Topics}} \textbf{\bibinfo{volume}{222}}, \bibinfo{pages}{1005--1016}
  (\bibinfo{year}{2013}).
\newblock \urlprefix\url{https://doi.org/10.1140/epjst/e2013-01902-4}.

\bibitem{Stojchevska_2011}
\bibinfo{author}{Stojchevska, L.} \emph{et~al.}
\newblock \bibinfo{title}{Mechanisms of nonthermal destruction of the
  superconducting state and melting of the charge-density-wave state by
  femtosecond laser pulses}.
\newblock \emph{\bibinfo{journal}{Phys. Rev. B}} \textbf{\bibinfo{volume}{84}},
  \bibinfo{pages}{180507} (\bibinfo{year}{2011}).
\newblock \urlprefix\url{https://link.aps.org/doi/10.1103/PhysRevB.84.180507}.

\bibitem{Pokharel_2022}
\bibinfo{author}{Pokharel, G.} \emph{et~al.}
\newblock \bibinfo{title}{Highly anisotropic magnetism in the vanadium-based
  kagome metal {Tb}{V}$_6${Sn}$_6$}.
\newblock \emph{\bibinfo{journal}{arXiv:2205.15559}}  (\bibinfo{year}{2022}).

\bibitem{PhysRevMaterials.6.104202}
\bibinfo{author}{Pokharel, G.} \emph{et~al.}
\newblock \bibinfo{title}{Highly anisotropic magnetism in the vanadium-based
  kagome metal {Tb}{V}$_6${Sn}$_6$}.
\newblock \emph{\bibinfo{journal}{Phys. Rev. Mater.}}
  \textbf{\bibinfo{volume}{6}}, \bibinfo{pages}{104202} (\bibinfo{year}{2022}).
\newblock
  \urlprefix\url{https://link.aps.org/doi/10.1103/PhysRevMaterials.6.104202}.

\bibitem{doi:10.1063/5.0005082}
\bibinfo{author}{Giannozzi, P.} \emph{et~al.}
\newblock \bibinfo{title}{Quantum espresso toward the exascale}.
\newblock \emph{\bibinfo{journal}{The Journal of Chemical Physics}}
  \textbf{\bibinfo{volume}{152}}, \bibinfo{pages}{154105}
  (\bibinfo{year}{2020}).
\newblock \urlprefix\url{https://doi.org/10.1063/5.0005082}.

\bibitem{Giannozzi_2009}
\bibinfo{author}{Giannozzi, P.} \emph{et~al.}
\newblock \bibinfo{title}{Quantum espresso: a modular and open-source software
  project for quantum simulations of materials}.
\newblock \emph{\bibinfo{journal}{Journal of Physics: Condensed Matter}}
  \textbf{\bibinfo{volume}{21}}, \bibinfo{pages}{395502}
  (\bibinfo{year}{2009}).
\newblock \urlprefix\url{https://dx.doi.org/10.1088/0953-8984/21/39/395502}.

\bibitem{Giannozzi_2017}
\bibinfo{author}{Giannozzi, P.} \emph{et~al.}
\newblock \bibinfo{title}{Advanced capabilities for materials modelling with
  quantum espresso}.
\newblock \emph{\bibinfo{journal}{Journal of Physics: Condensed Matter}}
  \textbf{\bibinfo{volume}{29}}, \bibinfo{pages}{465901}
  (\bibinfo{year}{2017}).
\newblock \urlprefix\url{https://dx.doi.org/10.1088/1361-648X/aa8f79}.

\bibitem{PhysRevLett.77.3865}
\bibinfo{author}{Perdew, J.~P.}, \bibinfo{author}{Burke, K.} \&
  \bibinfo{author}{Ernzerhof, M.}
\newblock \bibinfo{title}{Generalized gradient approximation made simple}.
\newblock \emph{\bibinfo{journal}{Phys. Rev. Lett.}}
  \textbf{\bibinfo{volume}{77}}, \bibinfo{pages}{3865--3868}
  (\bibinfo{year}{1996}).
\newblock \urlprefix\url{https://link.aps.org/doi/10.1103/PhysRevLett.77.3865}.

\bibitem{PhysRevB.88.085117}
\bibinfo{author}{Hamann, D.~R.}
\newblock \bibinfo{title}{Optimized norm-conserving vanderbilt
  pseudopotentials}.
\newblock \emph{\bibinfo{journal}{Phys. Rev. B}} \textbf{\bibinfo{volume}{88}},
  \bibinfo{pages}{085117} (\bibinfo{year}{2013}).
\newblock \urlprefix\url{https://link.aps.org/doi/10.1103/PhysRevB.88.085117}.

\bibitem{vaspkit}
\bibinfo{author}{Wang, V.}, \bibinfo{author}{Xu, N.}, \bibinfo{author}{Liu,
  J.-C.}, \bibinfo{author}{Tang, G.} \& \bibinfo{author}{Geng, W.-T.}
\newblock \bibinfo{title}{Vaspkit: A user-friendly interface facilitating
  high-throughput computing and analysis using vasp code}.
\newblock \emph{\bibinfo{journal}{Computer Physics Communications}}
  \textbf{\bibinfo{volume}{267}}, \bibinfo{pages}{108033}
  (\bibinfo{year}{2021}).
\newblock
  \urlprefix\url{https://www.sciencedirect.com/science/article/pii/S0010465521001454}.

\bibitem{vesta}
\bibinfo{author}{Momma, K.} \& \bibinfo{author}{Izumi, F.}
\newblock \bibinfo{title}{{{\it VESTA}: a three-dimensional visualization
  system for electronic and structural analysis}}.
\newblock \emph{\bibinfo{journal}{Journal of Applied Crystallography}}
  \textbf{\bibinfo{volume}{41}}, \bibinfo{pages}{653--658}
  (\bibinfo{year}{2008}).
\newblock \urlprefix\url{https://doi.org/10.1107/S0021889808012016}.

\bibitem{PhysRevB.105.165146}
\bibinfo{author}{Consiglio, A.} \emph{et~al.}
\newblock \bibinfo{title}{Van hove tuning of {$A$}{V}$_3${S}b$_5$ kagome metals
  under pressure and strain}.
\newblock \emph{\bibinfo{journal}{Phys. Rev. B}}
  \textbf{\bibinfo{volume}{105}}, \bibinfo{pages}{165146}
  (\bibinfo{year}{2022}).
\newblock \urlprefix\url{https://link.aps.org/doi/10.1103/PhysRevB.105.165146}.

\end{thebibliography}

\newpage

\newpage

\newpage

\newpage

\newpage

\section{Supplementary Information}

\section{ARPES and measurements without surface states}
The ARPES measurements have been performed across the transition temperature, i.e. 98~K, for ScV$_6$Sn$_6$ both with and without the present of surface states. While the surface states presence is guaranteed by a successful UHV cleave, in order to suppress them and to remain on the same sample's spot precautions are needed. Indeed, we noticed that by cleaving the samples both at 16~K and at 120~K (in the CDW phase and above it) the surfaces states are present. However, if one cleaves the samples at 16~K (or at 120~K) and varies the temperature up to 120~K (or down to 16~K) the surface states are killed, possibly to absorption onto the sample's following a temperature change. To a suppression of the surface states an overall reduction of the ARPES data quality was observed. In order to get high quality surface states, as in the main text, we cleaved the same sample one time at 16~K and one time at 120~K. Then, in order to detect the bulk electronic structure as in S1, we waited a sufficient time at the selected temperature until the surface states were suppressed by aging. This ensures a better overall quality of the ARPES data, compared to a forced aging by thermal process, and allows us to be sure to be on the same area of the sample with the light beam. We noticed also a variation in the aging time. For samples cleaved at 120~K, despite the initial lower quality, the surface states resisted for a longer time, i.e. 15-24 hours. For samples cleaved at 16~K, we noticed an initial better quality for the ARPES data, but a faster degradation time of about 4-6 hours. For the data in the main text, above the transition temperature, the surface states appear weaker. Such a weaker intensity recorded for the surface states above $T_{CDW}$ is not surprising given the more pronounced degradation expected for the highest temperature cleave. In this regard, we even noticed that by increasing slightly further the temperature, the surface states get completely suppressed. However, the electronic structure remains the same, with no changes in van-Hove singularities (red arrows in Fig.2 of the main text) and overall dispersion.

\section{ARPES and measurements along A-H and A-L directions}
As we showed in the main text, the DFT calculations captures very well the effects of the CDW on the electronic structure of ScV$_6$Sn$_6$. Along the high-symmetry directions $\Gamma$-M-K, we have the most favorable matrix elements and the strongest spectroscopic signal. In these conditions, the main changes are visible at the $\Gamma$ point as an increase in the energy separation of some states locates around -0.8 eV. Nevertheless, the DFT predicts the largest changes to occur in the $k_x$-$k_y$ plane located between two consecutive $\Gamma$-points. Here we used 65 eV photon energy (with both linear vertical and horizontal light polarizations) to capture such a plane and also the A-H and A-L directions, as shown in the Brillouin zone of Fig.1 of the main text. To note that despite the CDW-induced changes are supposed to be prominent here, the photoemission matrix elements reduces significantly the intensity of the electronic structure and, together with a rather significant $k_z$ broadening, the CDW changes are not easily detected. This is further challenged by the fact that in the low temperature phase, the spectral weight rapidly dimishes away from the gaps and still, the main intensity 'follow' the dispersion of the bands above $T_{CDW}$, as visible in the DFT of the main text. However, we will here show the results in these conditions.\\

First of all, we show the Fermi surface maps of the systems within the $k_x$-$k_y$ plane centred in the A point, along with the dispersion along the L-A-H-L, in figure 3 and 4, respectively. One can notice that apart from the better intensity and resolution gained in the low temperature phase and due to sharper features, overall there are no significant changes observed or at least, the changes are not big enough to justify them as effect of the CDW. To better see possible changes, we also report the very-high-resolution ARPES data along the A-H-L direction. Still, the data corroborate our previous conclusions.\\

\section{Tr-OS analysis and details}
In order to describe the incoherent part of the $\Delta R/R$ signal, we used a double-exponential function, $\Delta R/R \propto A_1e^{-t/\tau_1} + A_2e^{- t/\tau_2} + bkg$, where $A_1e^{-t/\tau_1}$ describes the fast decay process (first stage of the CDW) while $A_2e^{-t/\tau_2}$ describes the slower recovery process (second stage of the CDW). The $bkg$ term is used to take into account an extremely slow recovery process that can be described by a constant value in our time window.\\

In order to quantitatively analyze the oscillatory component of the $\Delta R/R$ signal, we considered the Fourier transform of the residuals obtained by considering the difference between the data and the double-exponential fit function described above. The complete fit function, resulting in the black lines of Fig.3 of the main text, has been obtained by adding a damped cosine oscillation, $A_{p}e^{-t/\tau_{p}}cos(w_{p}t+\phi_{p})$.\\

To give a rough estimate the induced heating by the pump, we assumed, reasonably, that the heat capacity of ScV$_6$Sn$_6$ is of the same order of magnitude of the ones reported for similar kagome systems (both binary and ternary). We do not expect that the temperature deviates more than 20 K upon pump application.

\section{Electronic and phononic band structures details}
In this section are reported complementary figures, supporting the statements and the results of the main text. In particular, we report the calculated electronic structure with and without CDW separately (Fig.  8), and the phonon density of states and the mechanism by which the phonon frequency has been estimated (Fig. 9). In addition (Fig. 10), we also show the data from time-resolved optical spectroscopy documenting the evolution of the fast component. Such evolution is reminiscent of the one observed in time-resolved photoelectron spectroscopy, and associated to the electronic degrees of freedom.

\newpage

\begin{figure*}
\centering
\includegraphics[width=1\textwidth,angle=0,clip=true]{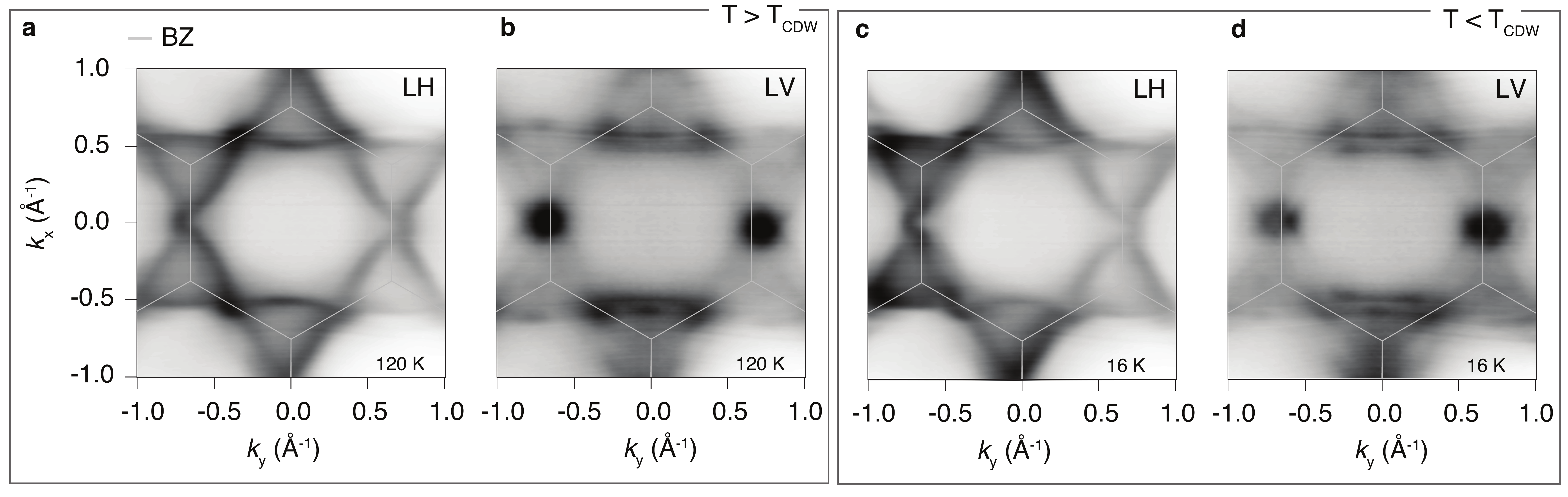}
\caption{{\bf Fermi surfaces of ScV$_6$Sn$_6$ kagome metal without surface states.} {\bf a} Linear horizontal and {\bf b} linear vertical polarization Fermi surface maps collected for ScV$_6$Sn$_6$ above the transition temperature. {\bf c} Linear horizontal and {\bf d} linear vertical polarization Fermi surface maps collected for ScV$_6$Sn$_6$ below the transition temperature. In order to be sure about the absence of surface states, the samples were left for a significant time in the same positions and same conditions (approximately 24 h) and they showed a natural aging effect, which resulted in the removal of the surface states manifold.} 
\label{S1}
\end{figure*}

\begin{figure*}
\centering
\includegraphics[width=1\textwidth,angle=0,clip=true]{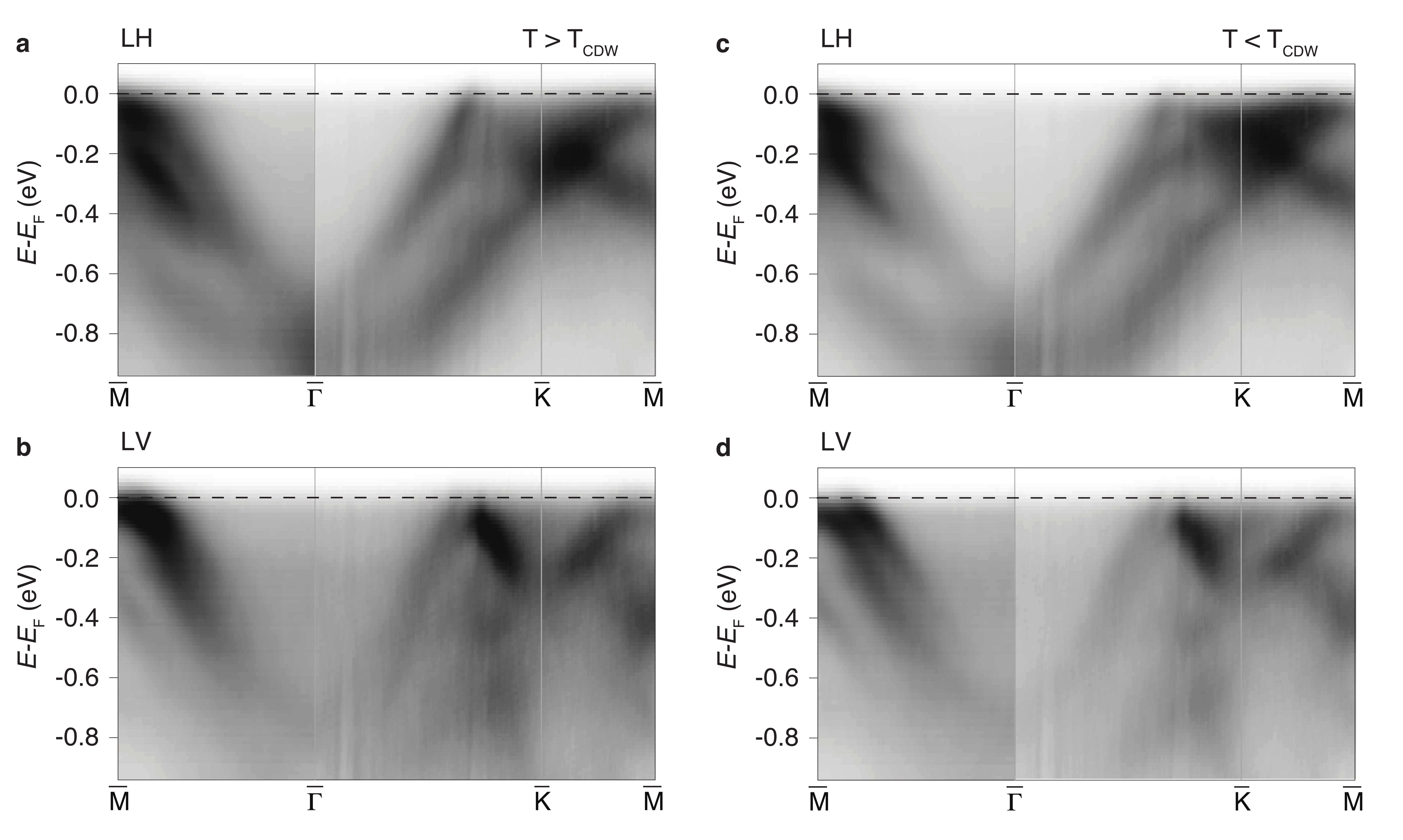}
\caption{{\bf Energy vs momentum dispersion of ScV$_6$Sn$_6$ kagome metal without surface states.} {\bf a} linear horizontal and {\bf b} vertical polarization spectra along the $\Gamma$-K-M direction of the BZ obtained for ScV$_6$Sn$_6$ after the surface states were suppressed as in S1, above $T_{CDW}$. {\bf c} linear horizontal and {\bf d} vertical polarization spectra along the $\Gamma$-K-M direction of the BZ obtained for ScV$_6$Sn$_6$ after the surface states were suppressed as in S1, below $T_{CDW}$.} 
\label{S2}
\end{figure*}

\begin{figure*}
\centering
\includegraphics[width=1\textwidth,angle=0,clip=true]{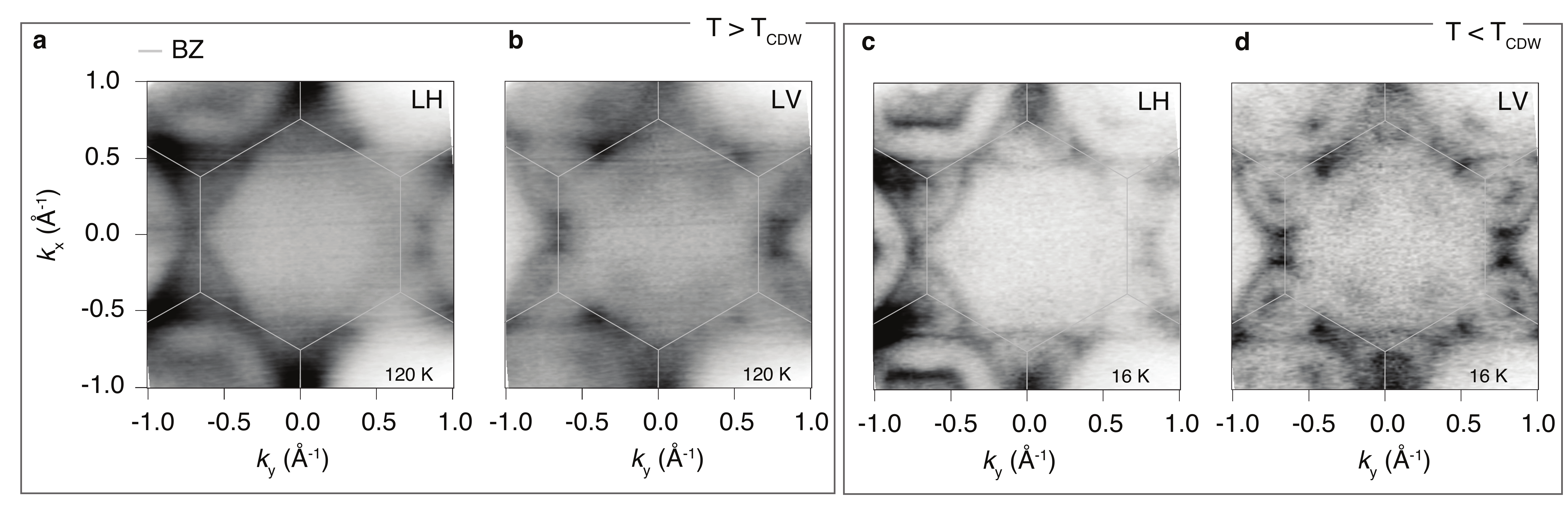}
\caption{{\bf Fermi surfaces of ScV$_6$Sn$_6$ kagome metal centred in the plane of the A point, i.e. 65 eV.} {\bf a} Linear horizontal and {\bf b} linear vertical polarization Fermi surface maps collected for ScV$_6$Sn$_6$ above the transition temperature. {\bf c} Linear horizontal and {\bf d} linear vertical polarization Fermi surface maps collected for ScV$_6$Sn$_6$ below the transition temperature. As one can see, the low temperature data appear sharper due to the reduced thermal broadening.} 
\label{S3}
\end{figure*}

\begin{figure*}
\centering
\includegraphics[width=1\textwidth,angle=0,clip=true]{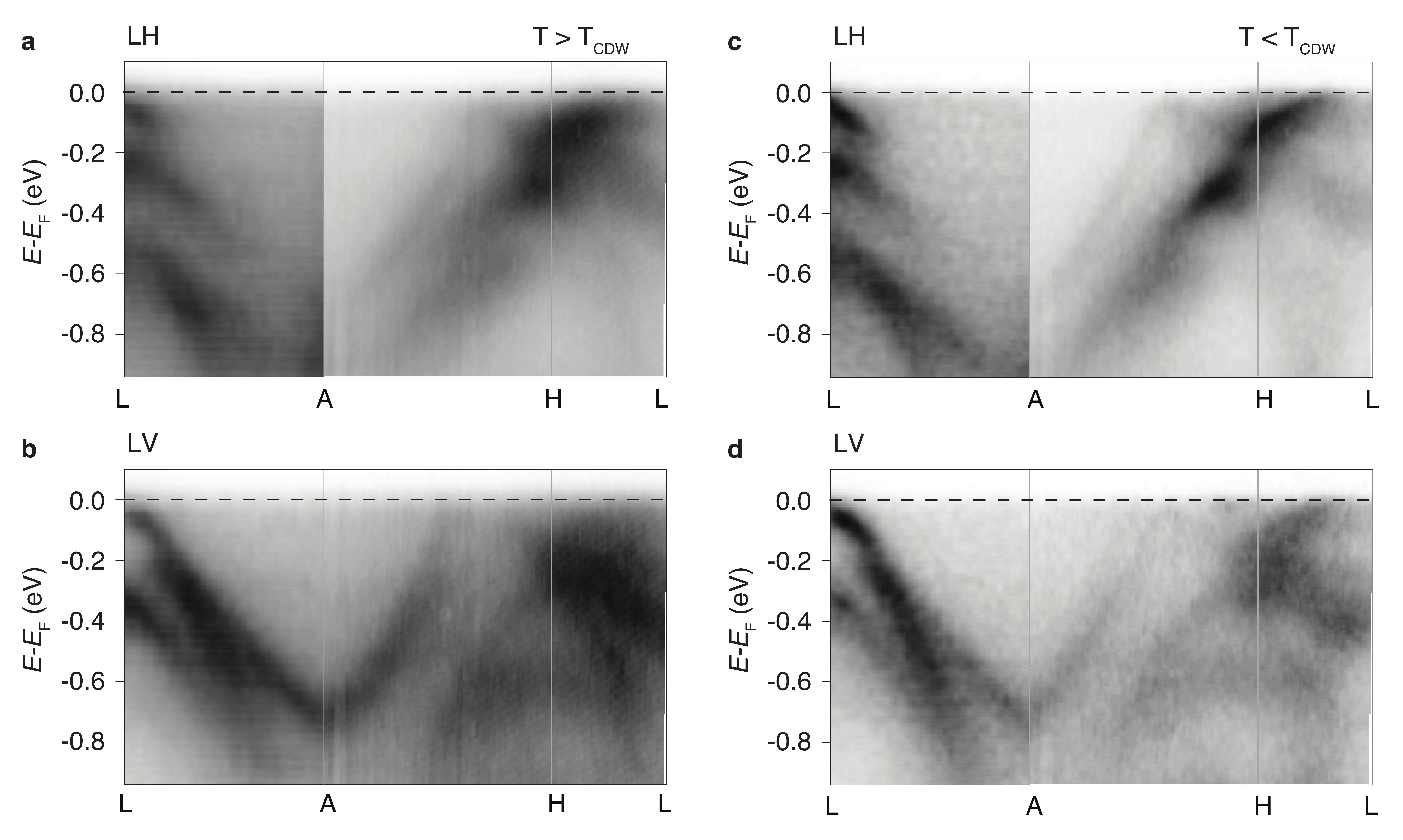}
\caption{{\bf Energy vs momentum dispersion of ScV$_6$Sn$_6$ kagome metal centred in the plane of the A point, i.e. 65 eV.} {\bf a} linear horizontal and {\bf b} vertical polarization spectra along the L-A-H-L direction of the BZ obtained for ScV$_6$Sn$_6$, above $T_{CDW}$. {\bf c} linear horizontal and {\bf d} vertical polarization spectra along the L-A-H-L direction of the BZ obtained for ScV$_6$Sn$_6$, below $T_{CDW}$.} 
\label{S4}
\end{figure*}

\begin{figure*}
\centering
\includegraphics[width=1\textwidth,angle=0,clip=true]{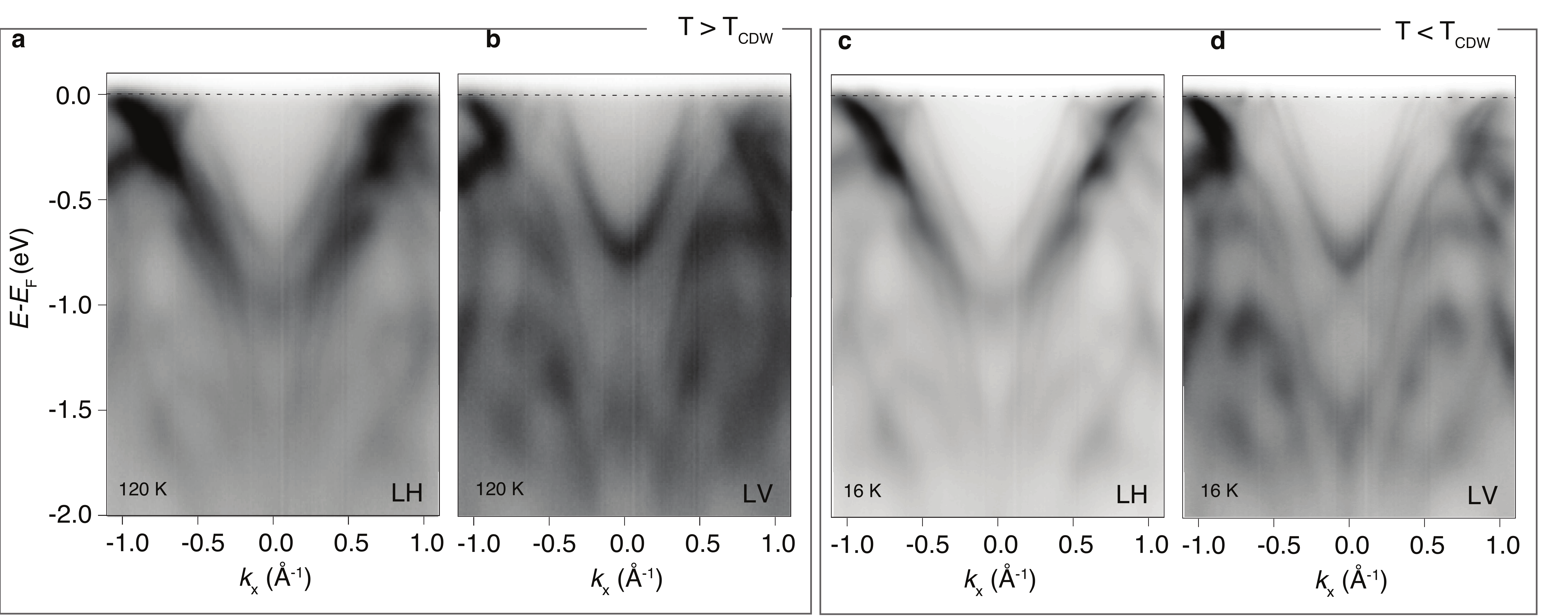}
\caption{{\bf High-resolution Energy vs momentum dispersion of ScV$_6$Sn$_6$ kagome metal along the A-H-L directions} {\bf a} linear horizontal and {\bf b} vertical polarization spectra along the A-H-L direction of the BZ obtained for ScV$_6$Sn$_6$, above $T_{CDW}$. {\bf c} linear horizontal and {\bf d} vertical polarization spectra along the A-H-L direction of the BZ obtained for ScV$_6$Sn$_6$, below $T_{CDW}$.} 
\label{S5}
\end{figure*}

\begin{figure*}
\centering
\includegraphics[width=0.8\textwidth,angle=0,clip=true]{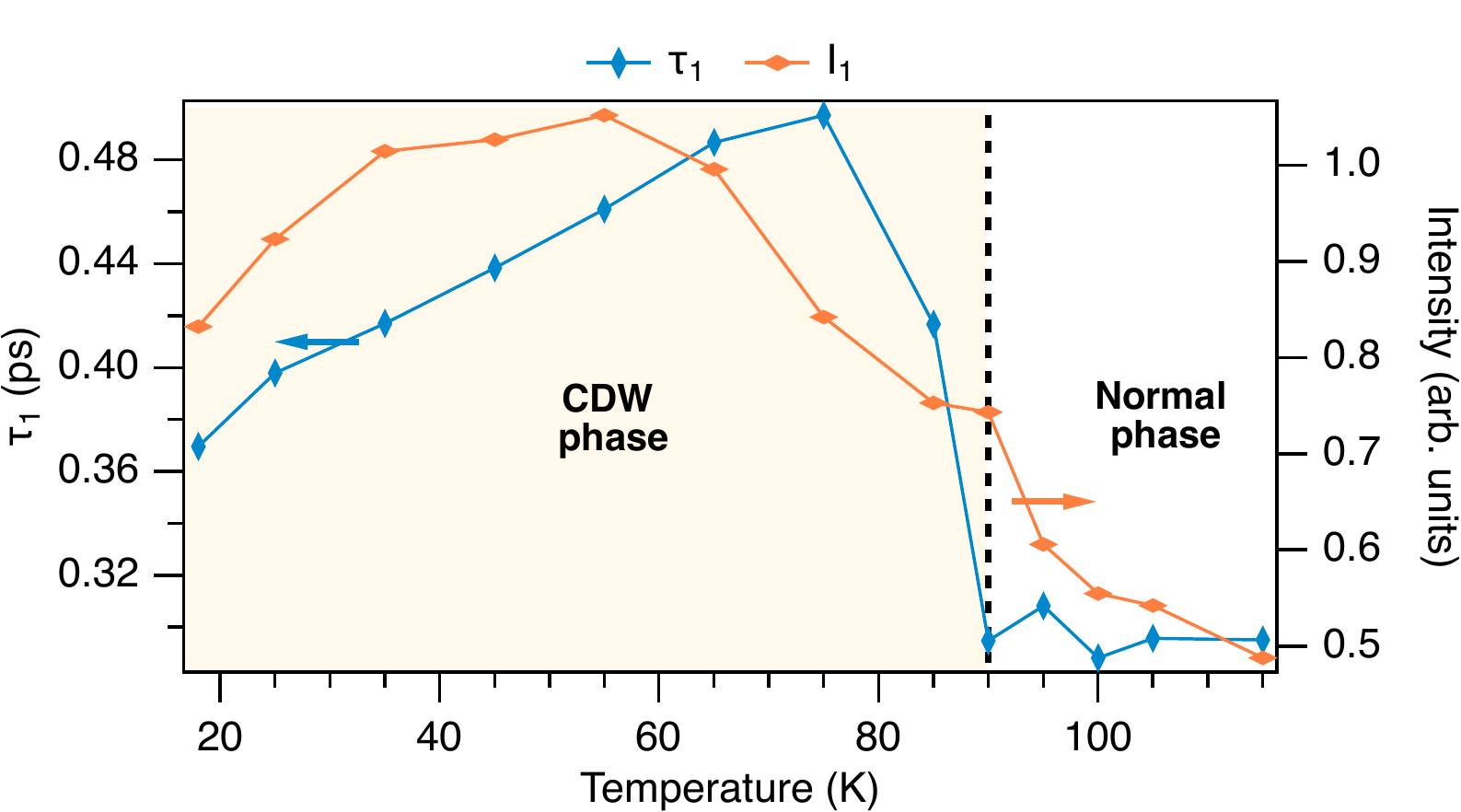}
\caption{{\bf Fast decay evolution as function of temperature.} Evolution of the lifetime ($\tau_{1}$, left axis) and of the amplitude (I$_1$, right axis) of the first fast decay as a function of the temperature. As one can see, contrary to the amplitude decay as function of fluence (Fig. 4 of the main taxt), which shows a saturation, here, I$_1$ has a very similar behaviour to the lifetime as function of temperature.} 
\label{S6}
\end{figure*}

\begin{figure*}
\centering
\includegraphics[width=0.72\textwidth,angle=0,clip=true]{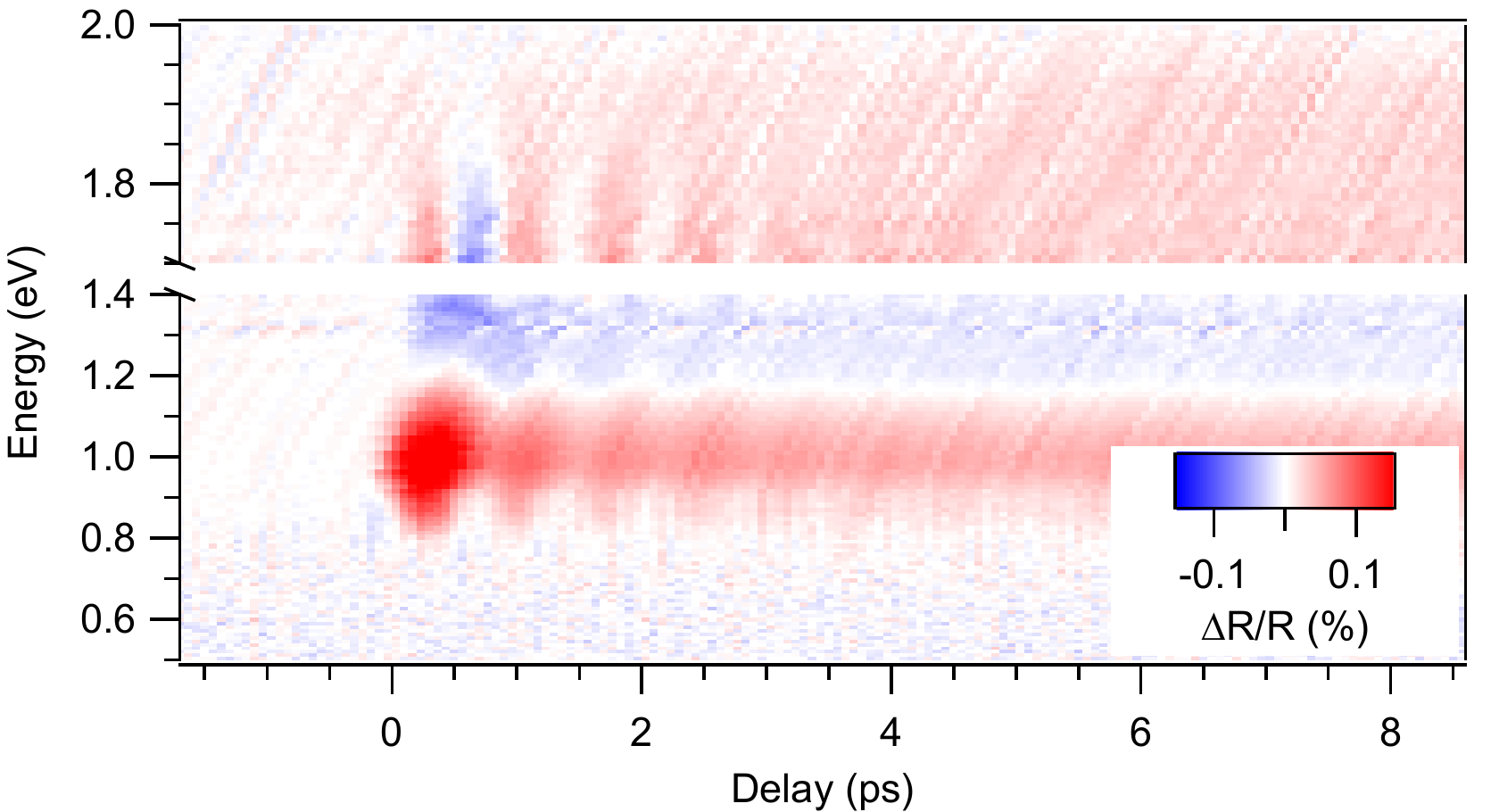}
\caption{{\bf Time- and frequency- resolved reflectivity.} The reflectivity is reported as energy versus time for samples kept at 30 K.} 
\label{S7}
\end{figure*}

\begin{figure*}
\centering
\includegraphics[width=1\textwidth,angle=0, trim={4cm  20.9cm  4.4cm  3.5cm},clip]{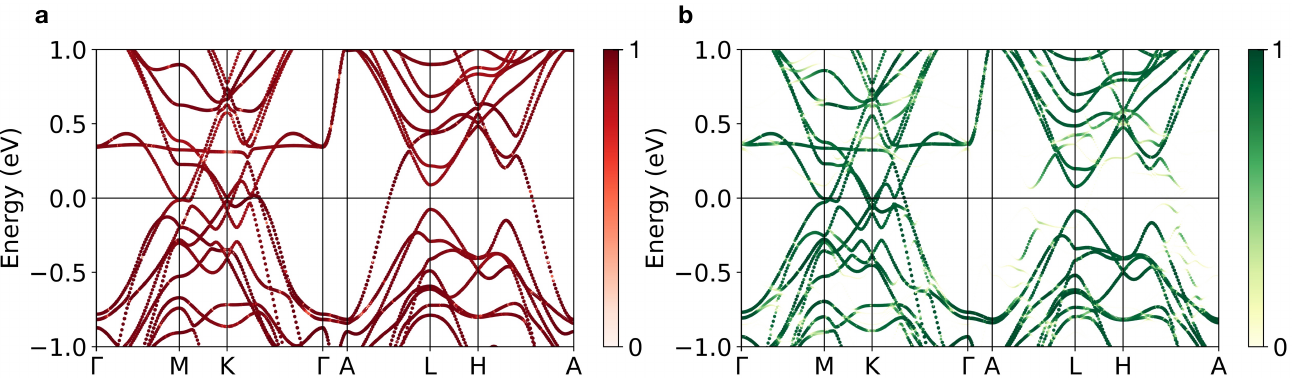}
\caption{{\bf Band structure unfolding of the pristine phase vs the CDW one} {\bf a} Electronic band structure of the pristine phase, with color-bar on the right side. The plot has been obtained unfolding the same superlattice of the CDW phase, without the lattice distortion. The bands have been shifted upward by 0.029 eV. {\bf b}  Electronic band structure of the CDW phase, with color-bar on the right side. Note how, in this case, a wider range of colors needs to be employed.} 
\label{S8}
\end{figure*}

\begin{figure*}
\centering
\includegraphics[width=1\textwidth,angle=0, trim={1.5cm  1.5cm  2cm  0.7cm},clip]{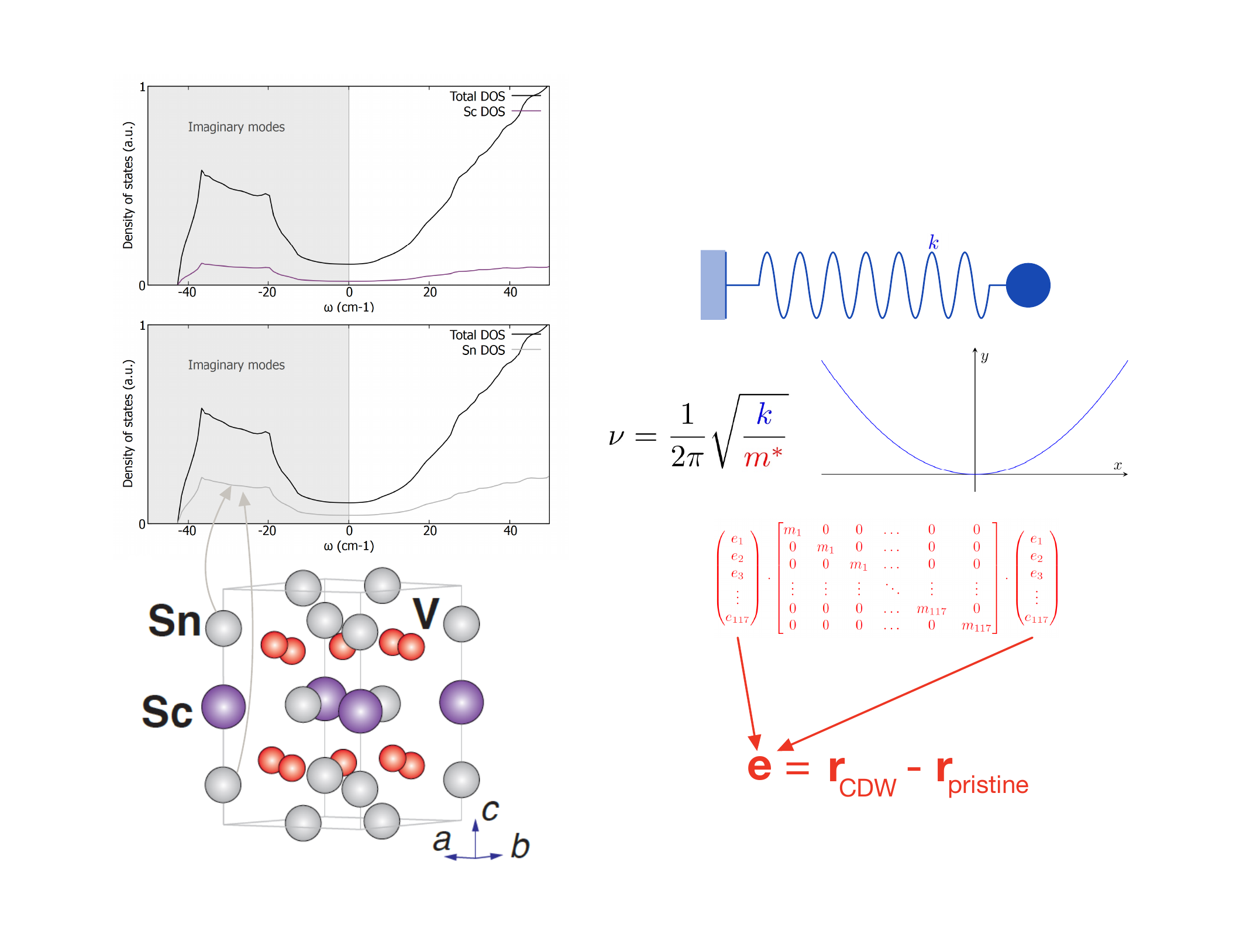}
\caption{{\bf Phonon and CDW theoretical details} {\bf a} Density of states of phonons in ScV6Sn6. The imaginary modes, indicating a dynamical instability, are only due to the Sc atom, and the Sn atoms with the same scandium's $(x, y)$ coordinates. Note that some atoms in the cell are represented more than once, because of the periodic boundary conditions. {\bf b}  Schematic sketch of the procedure followed to compute the frequency of the phonon mode ($\nu = 1.42$ THz). Red colors refer to the effective mass $m^*$, obtained from the mass tensor and the (normalized) displacement vector. Blue color refers to the spring constant, obtained by doubling the quadratic coefficient of the parabola. This parabola has been obtained via a quadratic fit, about one of the two minima of the double-well potential energy profile.} 
\label{S9}
\end{figure*}

\begin{figure*}
\centering
\includegraphics[width=0.5\textwidth,angle=0]{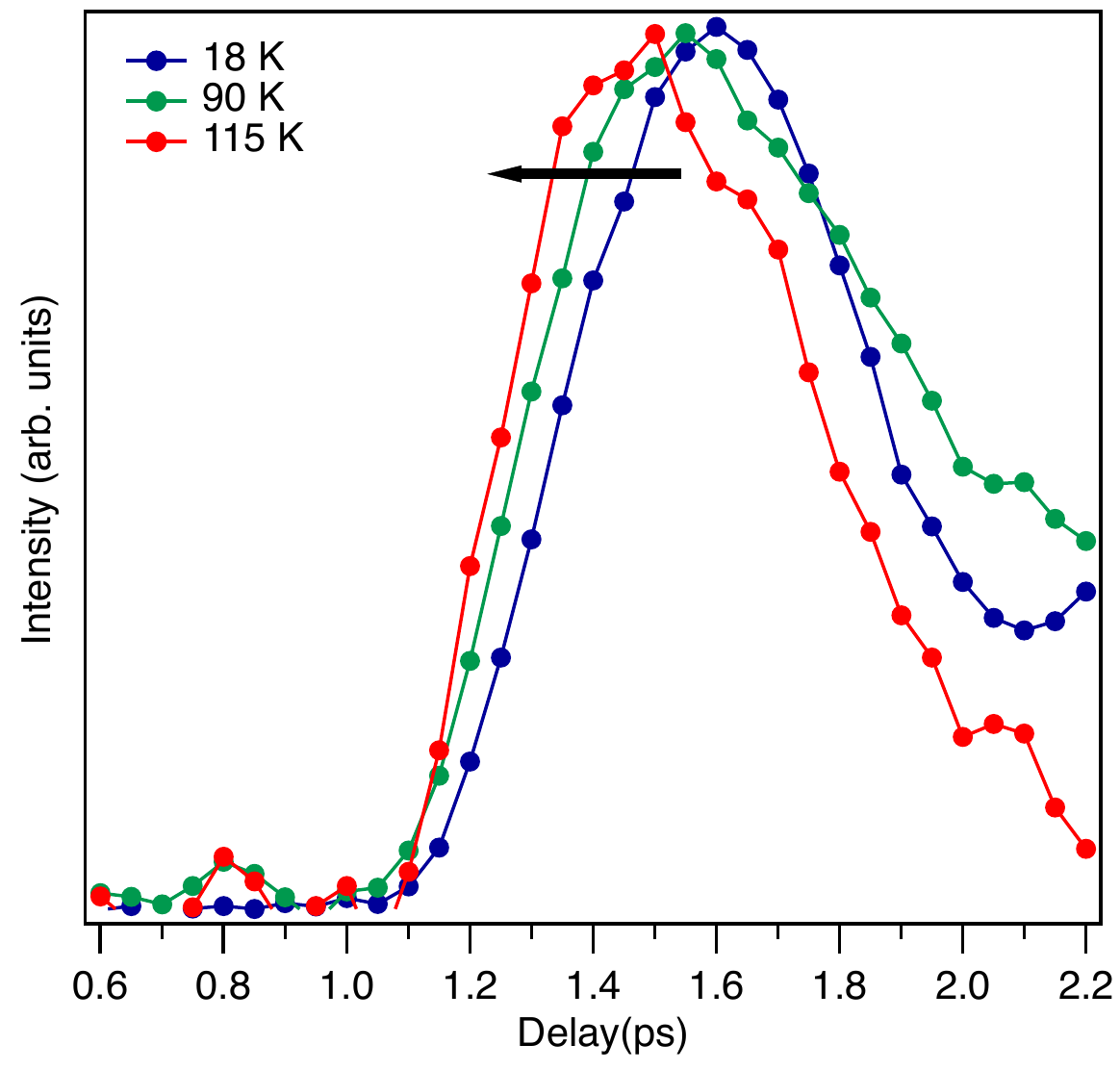}
\caption{{\bf Evolution of the rising edge as function of time and temperature.} Evolution of the rising edge of the $\Delta$R/R signal at the arrival of the pump pulse, for three selected temperatures (below $T_{CDW}$, at $T_{CDW}$ and above $T_{CDW}$). Above the critical temperature the maximum of the $\Delta$R/R  signal is reached ~150 fs faster than in the CDW phase.  The three traces have been normalized at their maximum intensity.} 
\label{rise}
\end{figure*}

\end{document}